\documentclass[lettersize,journal]{IEEEtran}

\usepackage{amsmath,amsfonts,amssymb}
\usepackage{array}
\usepackage{booktabs}
\usepackage{pifont}
\usepackage[dvipsnames]{xcolor}
\usepackage{float}
\usepackage{graphicx}
\usepackage{textcomp}
\usepackage{stfloats}
\usepackage{url}
\usepackage{cite}
\usepackage{makecell}
\usepackage{enumitem}
\usepackage[switch]{lineno}
\usepackage{tabularx}
\usepackage{microtype}
\usepackage{cuted}
\usepackage{wrapfig}

\DeclareGraphicsExtensions{.pdf,.png,.jpg,.jpeg}
\graphicspath{{figures/}{pictures/}{images/}{./}}
\definecolor{cohortgreen}{HTML}{a3b9b3}
\definecolor{cohortred}{HTML}{d7bbaa}
\definecolor{cohortblue}{HTML}{8198a2}
\definecolor{cohortpurple}{HTML}{b8a4ad}
\definecolor{cohortgray}{HTML}{c4c4c3}

\newcommand{\name}{{HPC-Vis}}

\newlength{\authorphotowidth}
\newlength{\authorbiosep}
\newlength{\authorbioskip}

\newcommand{\AuthorBio}[4][5]{%
  \par\noindent
  \begingroup
  \setlength{\intextsep}{0pt}%
  \setlength{\columnsep}{\authorbiosep}%
  \begin{wrapfigure}[#1]{l}{\authorphotowidth}
    \vspace{-0.1\baselineskip}
    \includegraphics[width=\authorphotowidth,clip,keepaspectratio]{#2}
  \end{wrapfigure}
  \textbf{#3} #4\par
  \endgroup
  \vspace{\authorbioskip}
}

\title{\name: A Visual Analytics System for Interactive Exploration of Historical Painter Cohorts}

\author{Yingping Yang,
Guangtao You,
Wenwen Li,
Jiayi Chen,
Yumeng Zhang,
Yuxin Lei,
Wei Zhang,
Jiazhou Chen,
and Wei Chen%
\thanks{Yingping Yang, Guangtao You, Wenwen Li, Jiayi Chen, Yumeng Zhang, Yuxin Lei, and Jiazhou Chen are with the College of Computer Science and Technology, Zhejiang University of Technology, Hangzhou, China. Jiazhou Chen is the corresponding author: cjz@zjut.edu.cn.}%
\thanks{Wei Zhang is with the School of Computer and Computing Science, Hangzhou City University, Hangzhou, China. Email: zw\_yixian@hzcu.edu.cn.}%
\thanks{Wei Chen is with the State Key Lab of CAD\&CG and the Laboratory of Art and Archaeology Image, Zhejiang University, Hangzhou, China. Email: chenvis@zju.edu.cn.}}

\begin{document}

\maketitle

\begin{abstract}
Painter cohort analysis has long been regarded as a key lens for studying how painting artistic styles develop and transmit across generations.
%Cohort analysis of historical painters is essential for understanding the evolution of painting art styles, 
Through a two-year collaboration with art historians, we identify key challenges in traditional painter cohort research: the unstructured characteristic of painter features, the entangled complexity of inheritance relationships, and the cognitively demanding nature of cohort definition and validation. 
To solve these challenges, we propose HPC-Vis, a visual analytics system for interactive exploration of historical painter cohorts. An improved cohort analytical workflow is designed to integrate structured feature construction, visualization-assisted exploration, algorithm-based recommendation, and unified cohort management. 
Based on this workflow, we develop three core computational modules: a multi-scale artistic feature construction method that leverages LLMs to extract and organize hierarchical style features from unstructured historical texts, an inheritance reconstruction algorithm that transforms the entangled multi-parent inheritance network into a clear hierarchical forest structure, and a recommendation model that identifies core features of the cohort and recommends cohort members via painter relevance assessment. 
To support smooth interactive exploration, we further design a set of novel visualizations with multidimensional collaboration, especially an inheriting mountain view inspired by traditional Chinese landscape paintings, and a foldable doughnut chart for hierarchical artistic style labels. HPC-Vis is evaluated and validated through case studies, user studies, and technical evaluations, demonstrating its effectiveness in supporting painter cohort exploration and in providing visual insights for art historical research.

\end{abstract}


\section{{Introduction}}
\label{sec:intro}

% 第一段：研究意义
Cohort analysis of historical painters studies groups of painters who share similar features, such as artistic styles, inheritance relationships, localities, social identities, etc. In Chinese art history, well-known painter cohorts such as the Wu School and the Zhe School, are often constituted by a founding master who establishes a distinctive style, a group of followers who share similar artistic approaches
%, and successive generations of inheritors who carry the tradition forward
~\cite{zhongguohuapailun,wangInheritanceRelationships2022}.
% In Chinese art history, well-known painter cohorts such as the Wu School and the Zhe School, are often constituted by a founding master who establishes a distinctive style, direct disciples who inherit the master's techniques through apprenticeship, and later followers who carry the tradition forward through stylistic imitation~\cite{zhongguohuapailun,wangInheritanceRelationships2022}. 
Painter cohorts have long been regarded as a key lens for studying how artistic styles develop and transmit across generations. Such cohort-level analysis provides insights beyond the study of individual painters, enabling a systematic understanding of the broader patterns and dynamics in art history. However, exploring and analyzing these cohorts is not straightforward.

% 第三段：画家群体的特殊挑战
Through a two-year close collaboration with four art historians, we identified a series of challenges that remain insufficiently addressed. \textbf{On the data side}, painter features are largely unstructured and multi-scale. Artistic styles are described in natural language scattered across multiple historical sources, and the inheritance network forms a complex multi-parent directed acyclic graph where lineages frequently cross and overlap (Fig.~\ref{fig:graph}), making it extremely difficult to organize and analyze. \textbf{On the analytical side}, the cohort definition and analysis process is heavily experience-dependent and cognitively demanding: researchers must manually formulate hypotheses, cross-reference multidimensional features to validate them, and maintain multiple evolving cohorts through a sustained and time-consuming exploration process.%, all without adequate computational or visual support.

% 第二段：现有方法——从自动聚类到可视分析
% In the broader literature, cohort analysis has been widely explored through computational methods.
Computational methods for cohort analysis have been widely explored in the literature.
A prevalent approach is to cluster individuals based on network structures and multidimensional similarities~\cite{wang2024classification,kang2023deep}. However, in the digital humanities domain, fully automatic clustering algorithms often fail to satisfy domain experts, as the nuanced criteria for clustering historical figures are difficult to formalize algorithmically and clustering results are difficult to understand and trust. To bridge this gap, visual analytics approaches have been proposed to integrate expert domain knowledges into the cohort exploration process through interactive interfaces, such as KP-Clustering for arbitrary social networks~\cite{pister2020integrating} and CohortVA for Chinese historical figures~\cite{zhang2022cohortva}. However, these methods are still limited when applied to painter cohort research: they treat relationships as flat attributes without dedicated support for visualizing complex inheritance structures, lack multi-scale feature exploration, and offer limited capabilities for sustained cohort management throughout a progressive exploration process.

% 第四段：我们的方案
To address these limitations, we propose \textbf{HPC-Vis}, a visual analytics system for interactive exploration of historical painter cohorts. An improved analytical workflow is first designed to integrate structured feature construction, visualization-assisted exploration, algorithm-based recommendation, and unified cohort management. For structured feature construction, we develop an LLM-based pipeline to automatically extract and organize multi-scale artistic style features from unstructured texts, and propose an inheritance reconstruction algorithm that transforms the entangled inheritance network into a clear hierarchical forest structure. For cohort definition and analysis, the system provides two complementary approaches: coordinated multi-dimensional visualizations help researchers discover patterns and validate hypotheses, including a novel \textit{Inheriting Mountain View} inspired by traditional Chinese landscape paintings that intuitively reveals lineage structures and a recommendation model to identify core features of the cohort and assess painter relevance, providing computational guidance for cohort verification, refinement, and expansion.

The contributions of this work can be summarized as follows:

% 第五段：贡献
\begin{itemize}
    \item An improved workflow for historical painter cohort research that supports the definition, validation, refinement, and expansion of painter cohorts through multi-dimensional feature construction, visual exploration, algorithmic recommendation, and unified cohort management.
    %\item An improved workflow for historical painter cohort research that integrates structured feature construction, visualization-assisted exploration, automatic recommendation, and unified cohort management.
    % . Through a two-year collaboration with domain experts, we identify the key challenges in traditional painter cohort research and propose an improved analytical workflow 
    \item A reconstruction and visualization method for inheritance relationships. It transforms the entangled multi-parent directed acyclic graph into a clear hierarchical inheritance forest, which is visualized by the \textit{Inheriting Mountain View} inspired by traditional Chinese landscape paintings.
    %, to intuitively reveal lineage structures and support interactive exploration.
    \item A visual analytics system, \textbf{HPC-Vis}, for interactive exploration and analysis of historical painter cohorts. Through case studies, user studies, and technical evaluations, we validated the effectiveness of the proposed approach.
\end{itemize}

\section{{Related Work}}
\label{sec:related}

%本节中，我们回顾了数字人文领域视觉分析、社交网络视觉分析以及社交网络计算等相关研究工作。
%We review visual analytics for humanities, cohort analysis, relationship visualization and graph reconstruction.

% 数字人文可视化
\subsection{Visual Analytics for Humanities}
Digital humanities is an interdisciplinary field that integrates computational technology into the humanities~\cite{khulusi2019interactive}. A notable direction is the collaboration with visual analytics, where various visualization methods have been proposed to support humanists in diverse tasks. Recent examples include catalog annotation~\cite{shao2024cataanno}, pottery motif evolution analysis~\cite{li2024pm}, book circulation research~\cite{guo2023liberroad}, painting color analysis~\cite{chen2024colornetvis}, painting provenance exploration~\cite{zhang2024scrolltimes}, calligraphic style comparison~\cite{calliva}, and porcelain history investigation~\cite{porcevis}.
% el2016visual, 

For historical figure research, several visual analytics systems have been developed to address different analytical challenges. TimeScape~\cite{timescape} proposed a multi-resolution timeline for exploring large-scale biographical data with seamless navigation between overview and detail. LSN-VA~\cite{lsnva} modeled dynamic social networks of ancient Chinese literati, supporting interactive exploration of their spatiotemporal evolution. Visual reasoning systems~\cite{uncertaintyfigures} have been developed to address uncertainty in spatiotemporal events of historical figures by building knowledge graphs to capture missing data and errors. Our work extends this line of research by focusing specifically on painter cohort analysis, where unique challenges arise from unstructured artistic style features, entangled inheritance networks, and the need for sustained iterative exploration.

\subsection{Cohort Analysis}

Cohort analysis aims to identify and study groups of individuals who are closely related and share common characteristics. Traditional approaches primarily rely on statistical models and data mining algorithms~\cite{zhao2014fluxflow}. 
Graph clustering and community detection methods~\cite{traag2019louvain,laenen2020higher,speidel2015community} have been widely used in network analysis, and other clustering algorithms have also been applied to cohort discovery in various domains~\cite{kang2023deep,wang2024classification,yu2025lowest}.
% General-purpose graph clustering methods~\cite{blondel2008fast} have been widely used for community detection in networks, and other clustering algorithms have also been applied to cohort discovery in various domains~\cite{kang2023deep,wang2024classification,yu2025lowest}. 
However, fully automatic methods often fail to satisfy domain experts, as they neither provide guidance for algorithm selection nor incorporate prior knowledge for result evaluation~\cite{pister2020integrating,buono2024big}.

To address these limitations, visual analytics approaches have been proposed to integrate human expertise into the cohort exploration process through interactive interfaces. PK-Clustering~\cite{pister2020integrating} captures prior knowledge of social scientists as incomplete clusters and ranks multiple clustering algorithms according to their knowledge matching. CareerLens~\cite{wang2021interactive} supports multi-level visual exploration of historical career mobility data. TopicBubbler~\cite{feng2023topicbubbler} enables cross-level topic exploration in social media data. TaxThemis~\cite{lin2020taxthemis} combines automated computation with interactive visualization for iterative analysis of tax-related groups. CohortVA~\cite{zhang2022cohortva} supports cohort analysis of historical figures through a knowledge-graph-based identification model built upon the China Biographical Database~\cite{cbdb}.

However, these systems have several limitations when applied to painter cohort research. First, they primarily treat relationships as flat attributes for clustering or filtering, without dedicated support for visualizing and reconstructing complex inheritance structures, which are the structural backbone of painter cohort analysis. Second, they lack support for multi-scale feature exploration, where researchers need to flexibly switch between coarse and fine-grained features during analysis. Third, they offer limited capabilities for sustained cohort management, such as maintaining, nesting, and comparing multiple evolving cohorts throughout a progressive exploration process. HPC-Vis is designed to address these limitations.

\subsection{Relationship Visualization and Graph Reconstruction}

Visualizing relationships among entities is a fundamental task in visual analytics. Common representations include node-link diagrams~\cite{Gephi,graphviz}, adjacency matrices~\cite{bezerianos2010geneaquilts,van2021simultaneous}, tree and hierarchical layouts~\cite{munzner2015visualization}, and Sankey or flow-based diagrams~\cite{wu2014opinionflow}. These representations have been widely applied to social networks~\cite{lsnva}, biological networks~\cite{fan2024visual}, and knowledge graphs~\cite{zhang2022graphical}. However, when the underlying graph is large and dense, direct visualization often results in severe visual clutter, failing to reveal meaningful structures.

To improve readability and analytical efficiency, various graph reconstruction and abstraction strategies have been proposed~\cite{chen2018structure}. GraphInterpreter~\cite{lin2024graphinterpreter} leverages topic models to extract latent structural patterns from dynamic networks. Zhou et al.~\cite{zhou2020context} propose a context-aware sampling approach based on graph representation learning to abstract large-scale networks while preserving critical structures. In application scenarios, graph reconstruction has been applied to medical causal analysis~\cite{fan2024visual} and urban metro network exploration~\cite{chen2024city}, demonstrating adaptability across domains~\cite{xiang2025scalable}. % basole2018ecoxight

\begin{figure}[!t]
\centering
\includegraphics[width=0.9\linewidth]{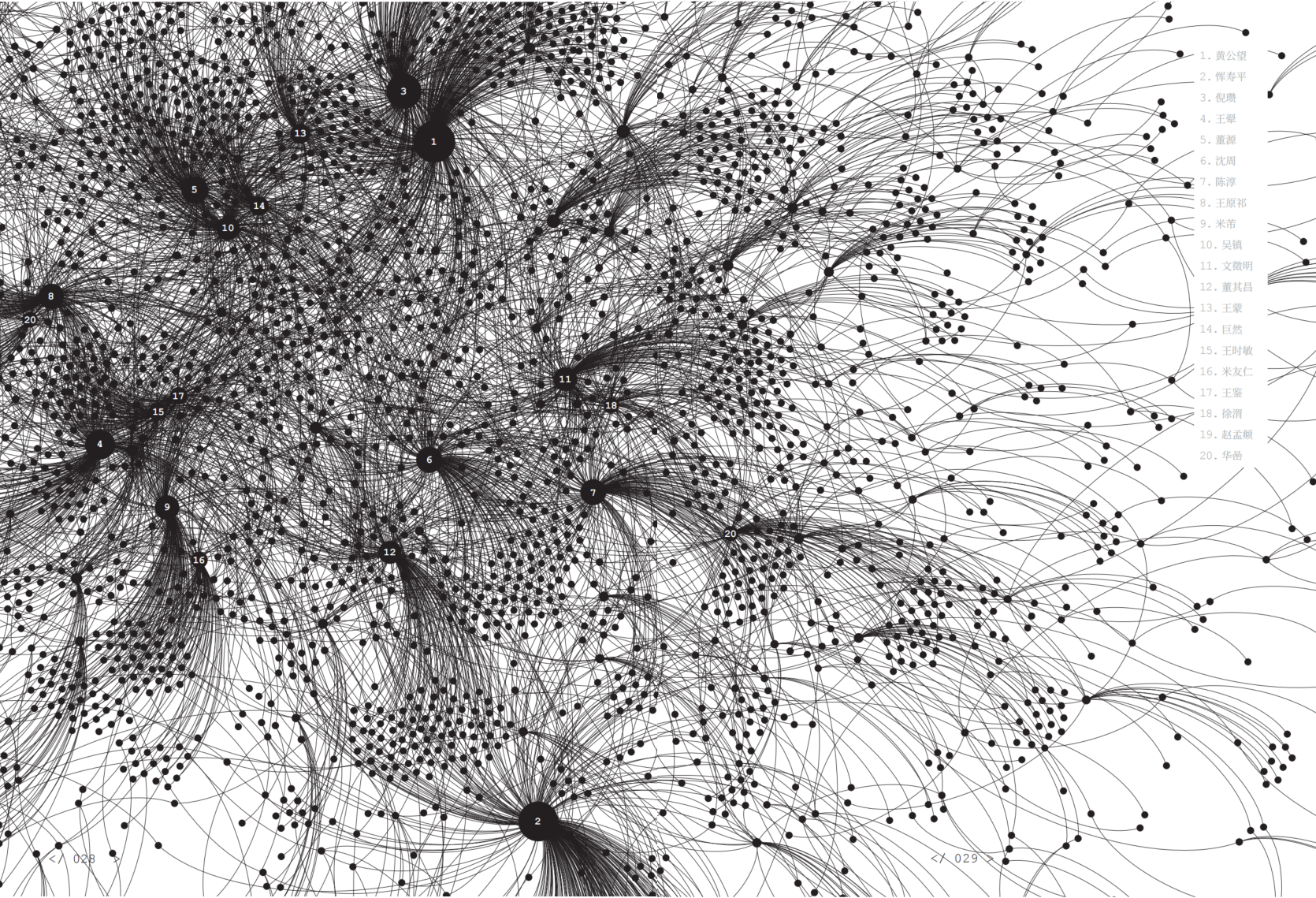}
\caption{
%Gelphi软件绘制的画家师承师法关系
A node-link visualization of inheritance relationships among Chinese historical painters from the book \textit{A Data Spectrum of Inheritance Relationships of Chinese Painters}~\cite{wang2022}, rendered using Gephi~\cite{Gephi}. The dense and intertwined structure illustrates the complexity of the raw inheritance network, making it difficult to identify painter cohorts.
}
\label{fig:graph}
\end{figure}

However, relationship visualization and reconstruction for historical painter inheritance networks remain underexplored. In such networks, painters often inherit from multiple predecessors, and lineages extend continuously across many generations, causing frequent crossing and overlapping among different lineages (Fig.~\ref{fig:graph}). Applying general graph visualization and reconstruction methods to such data often yields cluttered layouts that fail to reveal clear lineage structures. To address this, we propose a dedicated three-stage reconstruction algorithm that transforms the entangled inheritance network into a hierarchical forest structure with clear inheritance chains, and design the Inheriting Mountain View to intuitively visualize the reconstructed lineage structures.

\begin{figure*}[!t]
    \setlength{\belowcaptionskip}{-0.3cm}
    \centering 
    \includegraphics[width=0.95\textwidth]{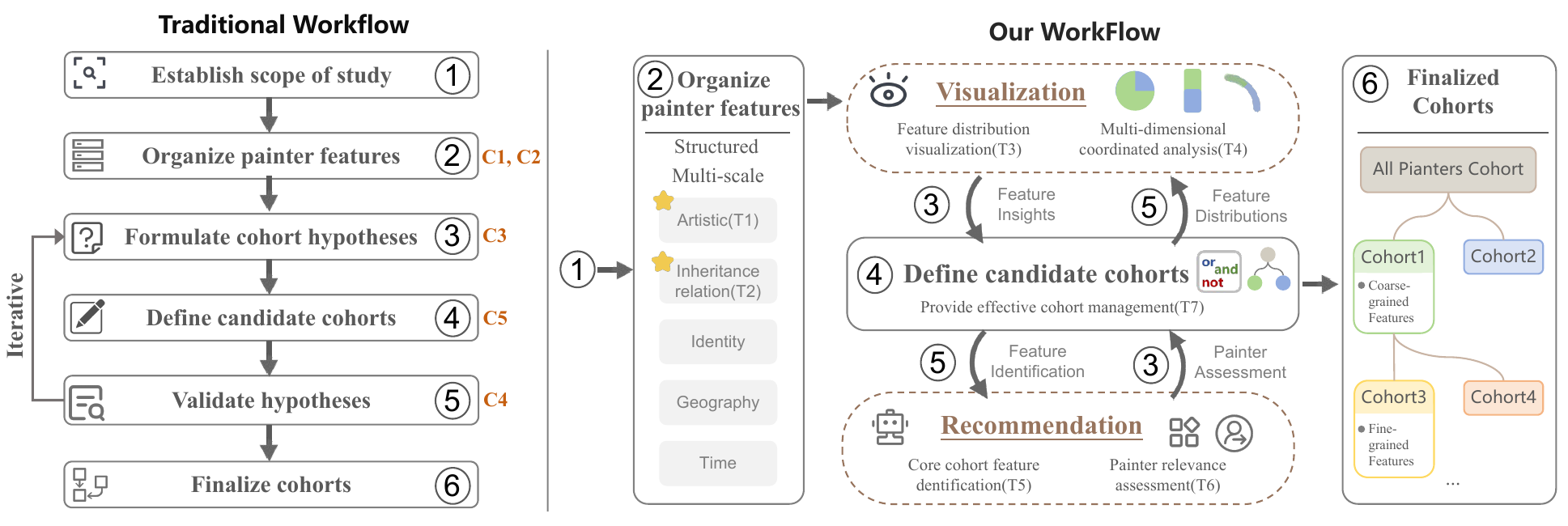}
    \caption{The traditional workflow (left) and our improved workflow (right) for painter cohort research.
   }
    \label{fig:Workflow}
\end{figure*}

\section{Background}

% 在过去两年中，我们与三位在画家群体研究领域经验丰富的历史学家密切合作。我们的合作贯穿了这项工作的各个阶段，从识别挑战到系统评估。在本节中，我们介绍我们的数据集，然后描述传统工作流程。
We have worked closely with four experienced historians in the field of painter cohort research over the past two years. Our collaboration spans all stages of this work. %In this section, we first introduce our dataset, then describe the traditional workflow and summarize important challenges.

\subsection{Data Description}
\label{sec:data-description}
%我们的数据来源于大型中国美术家数据库"云艺象"。该数据库通过对历代重要画论、品录及近代美术辞典等文献中记载的师承师法关系进行采集与处理，收录了从黄帝至民国初年三万多位书画家、篆刻家及鉴藏家的详细数据。对于每个画家，数据内容主要包括以下三个方面：
% 基本信息：包括字、号、性别、出生地、所属朝代、生卒年份、职位与头衔等数据；
% 艺术特点描述：涵盖多来源的文本描述，重点描述画家的艺术风格；
% 师承关系：记录包括师承师法、弟子后学、家族关系与友人交往等多类人际关联，其中师承关系为本研究的重点分析对象。
%我们依据《中国历代画家图表》所载名录，从“云艺象”数据库中提取了自汉代至清代的2212位画家数据，以支持后续的分析。
This study uses a large-scale database of Chinese artists from ``Artyx Cloud''\footnote{https://www.artyx.cn/}. It is constructed by manually extracting and organizing artist-related information from important painting theories, critical commentaries, and art dictionaries across most Chinese dynasties. It contains more than 30,000 artists from the Yellow Emperor period to the early Republic of China, including painters, calligraphers, engravers, and collectors. 
2212 painters are selected from the Artyx Cloud database based on the painter list in the book ``Chinese Painters Chart''\cite{ye2009chinese}. They are well-known Chinese historical artists from the Han Dynasty to the Qing Dynasty. Each of them has the following data:
\begin{itemize}[noitemsep, topsep=0pt, leftmargin=1em, labelindent=0pt, labelsep=0.5em]
\item \textit{Basic information:} name, surname, gender, place of birth, dynasty, year of birth and death, social identity, and title;
\item \textit{Artistic style descriptions:} multi-sourced textual descriptions characterizing each painter's artistic features. 
\item \textit{Inheritance relationship:} recorded associations of artistic lineage, including direct master-apprentice transmission and stylistic imitation by later generations.
\end{itemize}

% 传统工作流
\subsection{Traditional Workflow}
By interviewing domain experts in painter cohort research, we summarized their traditional workflow. % and summarize the challenges they are facing during their research practices.
% 如图2左侧，历史研究者证实了画家群体的传统研究流程。
% 研究者首先基于研究问题确定分析范围（例如特定地域、时间阶段的画家群体）。在此基础上，收集相关的画家数据，记录画家的多维特征，包括艺术风格描述、师承关系、身份背景、所在地以及时代信息等。这些数据通常以表格、文本摘录或研究笔记的形式进行记录，并辅以简单的关系图（如节点—连接图）来表达画家之间的关联。
As shown on the left of Fig.~\ref{fig:Workflow}, researchers begin by \textit{establishing the scope of study} according to their research question, such as painters from a particular region or period. They then \textit{organize painter features} by collecting multidimensional information, primarily including inheritance relationships and artistic styles, along with social identities, geography, dynasty, etc. These features are typically organized in tables, text excerpts, or research notes, supplemented with simple relational diagrams (e.g., node-link graphs) to depict connections among painters.

% 在完成数据整理后，研究者结合自身的领域经验，并基于对数据的初步观察，提出若干关于潜在画家群体的假设。随后，依据假设定义一个或多个候选群体。研究者进一步对定义群体的特征一致性进行分析，以检验假设的合理性。
Once the data have been compiled, researchers draw on their domain expertise and preliminary observations to \textit{formulate cohort hypotheses}. They then \textit{define candidate cohorts} based on these hypotheses and \textit{validate the hypotheses} by examining whether cohort members exhibit consistent features across the relevant dimensions.
% 在这一过程中，分析结果往往会带来新的启发，从而促使研究者修正已有假设或提出新的假设，并进入下一轮分析。通过多轮迭代，研究者逐步深化对研究问题的理解，最终形成相对稳定且具有解释力的画家群体定义。
This process is inherently iterative: analytical findings frequently spark new insights that lead researchers to refine existing hypotheses or formulate alternative ones, initiating a new round of analysis. Over successive iterations, researchers progressively sharpen their understanding of the research question and ultimately \textit{finalize cohort definitions} that are both stable and historically interpretable.

\subsection{Challenges}
% 尽管这一流程在长期实践中被广泛采用，但在实际操作中仍面临如下的挑战：
% Although the traditional workflow has been widely adopted over long-term practices, it still faces several challenges:
The traditional workflow faces the following challenges:

\textbf{C1: Painter features are unstructured and multi-scale.}
Painter features are mostly scattered across ancient books, scholarly literature, and archives in unstructured natural language. Artistic style features, for example, require manual integration from multiple sources. Cohort analysis also requires multi-scale reasoning, such as moving from figure painting to its sub-category, court-lady painting. Without a unified multi-scale organization, researchers must reorganize features whenever the analytical granularity changes.

% Painter features typically originate from ancient books, scholarly literature, and archived records, and are largely unstructured. Artistic style features, for example, are expressed in natural language and scattered across multiple sources, requiring researchers to manually integrate them. Furthermore, cohort analysis often demands reasoning at multiple scales: a researcher may initially study the cohort of painters proficient in figure painting, and later narrow the focus to those specializing in its sub-category, court-lady painting. The traditional workflow lacks a unified multi-scale organization of features, forcing researchers to re-organize painter features whenever they switch analytical granularity.

\textbf{C2: Inheritance relationships are dense and entangled.}
Inheritance relationships are the backbone of painter cohort analysis but difficult to organize. The raw inheritance network is a large multi-parent DAG where lineages frequently cross and overlap (Fig.~\ref{fig:graph}). Researchers must manually determine shared lineages, dominant inheritance paths, and inter-lineage relations, often spending days or weeks on a single lineage.

% Inheritance relationships are the backbone of painter cohort analysis, yet also very difficult to organize. The raw inheritance network is a large multi-parent directed acyclic graph in which lineages frequently cross and overlap, resulting in an extremely dense and tangled structure (Fig.~\ref{fig:graph}). From such a structure, it is extremely difficult for researchers to manually identify which painters belong to the same lineage, which inheritance paths are dominant, and how different lineages relate to one another. In practice, collaborating historians often spend days or even weeks tracing a single lineage, and the resulting structures are still hard to verify. A clear, structured representation of inheritance relationships is therefore urgently needed.

% C3 群体定义依赖经验。群体的定义依靠研究者提出合理的群体假设，这对研究者提出了较高的领域知识和研究经验的要求，如果不能提出合理的群体假设，研究者需要通过大量的迭代才能完善群体，整体效率较低。
% C3: Cohort hypothesis formulation relies heavily on experience.
\textbf{C3: Cohort definition relies heavily on experience.} Cohort definition depends on reasonable hypotheses, requiring extensive domain knowledge and research experience. If reasonable hypotheses cannot be proposed, researchers must repeatedly iterate to refine cohorts, reducing overall efficiency.

% The definition of cohorts depends on researchers proposing reasonable hypotheses, which demands extensive domain knowledge and research experience. If reasonable hypotheses cannot be proposed, researchers must iterate extensively to refine the cohorts, resulting in low overall efficiency.

% C4 群体特征分析不直观。对定义的一个或多个候选群体，研究者需要在群体内进行多维度的特征一致性分析，以及群体之间对比分析共性与差异。这一分析过程主要依赖研究者在不同的特征维度之间进行反复对照和归纳，这一过程不直观，具有较高的认知负担。
 \textbf{C4: Cohort validation across dimensions is cognitively demanding.} After defining candidate cohorts, researchers must validate them by checking feature consistency within cohorts and comparing commonalities and differences across cohorts. This requires repeated cross-referencing among scattered materials, which is cognitively demanding and error-prone.

% Once candidate cohorts have been defined, researchers need to validate their hypotheses by examining cohort members across multiple feature dimensions: assessing whether members within a cohort share consistent features, and identifying commonalities and differences across cohorts. This process relies heavily on repeatedly cross-referencing and synthesizing information from scattered tables, notes, and diagrams, which is cognitively demanding and error-prone.

% C5 持续迭代分析很难维护。画家群体研究是一个持续深入的过程，围绕一个研究问题，研究者往往会逐步构建出多个画家群体，包括具有对比意义的群体，以及在一个群体内部进一步划分更细分的子群体。然而，在传统工作流中，这些群体通常以分散的笔记形式存在，缺乏统一的组织与管理工具。研究者需要在脑中维护多个群体及其分析上下文，这不仅增加了认知负担，也使得系统性的持续探索变得困难。
\textbf{C5: Iterative cohort exploration is difficult to maintain.} Painter cohort analysis is a progressive process with a number of interactive explorations. Around a research question, researchers often construct multiple cohorts over time, including cohorts for comparative analysis and sub-cohorts within a larger cohort. However, in the traditional workflow, researchers have to mentally track the evolving relationships among them with scattered notes and documents, which not only increases cognitive load but also makes iterative exploration difficult.

\section{REQUIREMENTS AND NEW WORKFLOW}
\label{sec:requirement}
This section presents the design requirements and tasks (in A) distilled from the above challenges, and introduces our improved workflow (in B) against the traditional one.

% 任务分析
\subsection{Requirement \& Task Analysis}

% 根据上述挑战，我们总结了系统的设计需求和任务。
%Based on the challenges from domain experts, we summarize the design requirements and analytical tasks below.

\subsubsection{\textbf{Construct structured painter features}} To address the unstructured and multi-scale nature of painter data (C1) and the entangled inheritance relationships (C2), the system needs to convert raw data into structured, multi-granularity features suitable for cohort analysis. For our dataset (Section~\ref{sec:data-description}), it primarily produces two technical tasks: %artistic style features and inheritance relationships. Other dimensions are already structured and can be used directly after Simple processing.

%\hangindent=2.5em 
\textbf{T1: Artistic style feature extraction and structural organization.} The system should automatically extract artistic style features from unstructured texts and reconstruct a hierarchical structure to support analysis at multiple granularities.

%\hangindent=2.5em 
\textbf{T2: Inheritance relationship reconstruction.} The entangled inheritance network should be converted into a clear, structured representation that enables researchers to identify lineage groupings, trace dominant inheritance paths, and understand relationships among different lineages.

% 可视化协助群体定义与分析。针对群体定义依赖经验（C2）以及群体特征分析不直观（C3）的问题，系统需要提供有效的可视化协助研究者进行群体定义与分析。可视化能够将杂乱的画家特征转化为直观的视觉表达，协助研究者理解特征分布并进行合理的群体假设，同时通过多维协同的群体特征可视呈现，让群体验证变得轻松，加快群体分析迭代。
\subsubsection{\textbf{Visualization-assisted cohort definition and analysis}}
To address the experience reliance in cohort analysis (C3) and the non-intuitive nature of cohort validation (C4), the system should provide intuitive visualizations of painter feature distributions to support hypothesis formulation and validation, and enable coordinated analysis across multiple feature dimensions.

% T3：特征分布可视化。以直观的可视化形式呈现画家的多维特征分布，帮助用户快速理解特征分布与潜在模式，从而形成群体洞察并提出合理的群体假设；对于已构建的群体，能够突出显示群体相关的特征分布，以支持群体特征的一致性验证以及群体间的对比分析；支持多尺度特征的灵活展示与切换，以满足研究者在不同粒度下的分析需求。
%\hangindent=2.5em 
\textbf{T3: Feature distribution visualization.} The system should present painter feature distributions in intuitive visual formats to help users quickly identify underlying patterns and formulate reasonable cohort hypotheses. For defined cohorts, it should highlight cohort feature distributions to support intra-cohort consistency verification and inter-cohort comparative analysis, and support flexible display and switching of multi-scale features for analysis needs at different granularities.

% T4: 多维度协同分析。通过多视图协同的方式，使研究者能够在多个视角下交叉观察与验证群体特征，从而获得更全面的分析理解。
%\hangindent=2.5em 
\textbf{T4: Multi-dimensional coordinated analysis.} The system should allow researchers to define and adjust cohorts from any feature dimension, with cohort distributions automatically reflected across all other dimensions for comprehensive cross-dimensional examination.
 
% 推荐算法协助群体定义与分析。针对群体定义和验证中的挑战（C2，C3），在可视化支持之外，系统还需要引入自动算法辅助群体构建与分析。尽管可视化能够支持数据观察，但研究者仍需要在不同维度之间进行反复对比与归纳。因此，系统需通过算法自动分析群体特征分布，提供核心特征与相关画家群体推荐，从而在“观察”的基础上进一步提供“推断”与“引导”，辅助用户进行群体定义、验证与扩展，加快分析迭代过程。
\subsubsection{\textbf{Algorithm-assisted cohort definition and analysis}}
Beyond visualization, the system should incorporate algorithmic support for cohort definition and analysis (C3, C4). Although visualizations reveal feature distributions, manually deducing cohort features and verifying memberships remain cognitively demanding. Therefore, the system should automatically abstract shared cohort features and quantify painter relevance.

% Beyond visualization, the system should incorporate algorithmic support to assist cohort definition and analysis (C3, C4). While visualizations effectively reveal feature distributions, manually deducing a cohort's  features and verifying individual memberships across multidimensional data remain cognitively demanding. Therefore, the system should automatically abstract the shared features of a cohort and quantify the relevance of individual painters to these features.
% By providing algorithmic guidance on top of visual observation, the system can significantly accelerate the overall research and analytical process.

% Beyond visualization, the system should also incorporate algorithmic support to assist cohort definition and analysis (C3, C4). While visualization enables researchers to observe feature distributions, they still need to manually compare and synthesize across dimensions to identify representative features and discover relevant painters. The system should therefore provide automatic identification of core cohort features and recommendation of similar painters, offering computational inference and guidance on top of visual observation to assist cohort definition, verification, and expansion.

% T5：推荐群体核心特征。基于已定义的候选群体，自动识别并推荐具有代表性和一致性的特征，以支持群体的解释与假设验证。
%\hangindent=2.5em 
\textbf{T5: Core cohort feature identification.} 
The system should automatically identify the core features of a candidate cohort to support cohort interpretation and hypothesis verification.

% T6：推荐特征相似的群体。基于推荐的群体核心特征或用户指定特征，自动推荐与目标群体在关键特征上相似的画家群体，支持用户快速进行群体定义与扩展。
%\hangindent=2.5em 
\textbf{T6: Painter relevance assessment.} The system should evaluate the similarity of painters based on core features. It enables researchers to verify the feature consistency of internal cohort members and discover relevant external candidates, inspiring new hypotheses and supporting rapid cohort expansion.

\begin{figure*}[!t]
    \setlength{\belowcaptionskip}{-0.3cm}
    \centering 
    \includegraphics[width=0.95\textwidth]{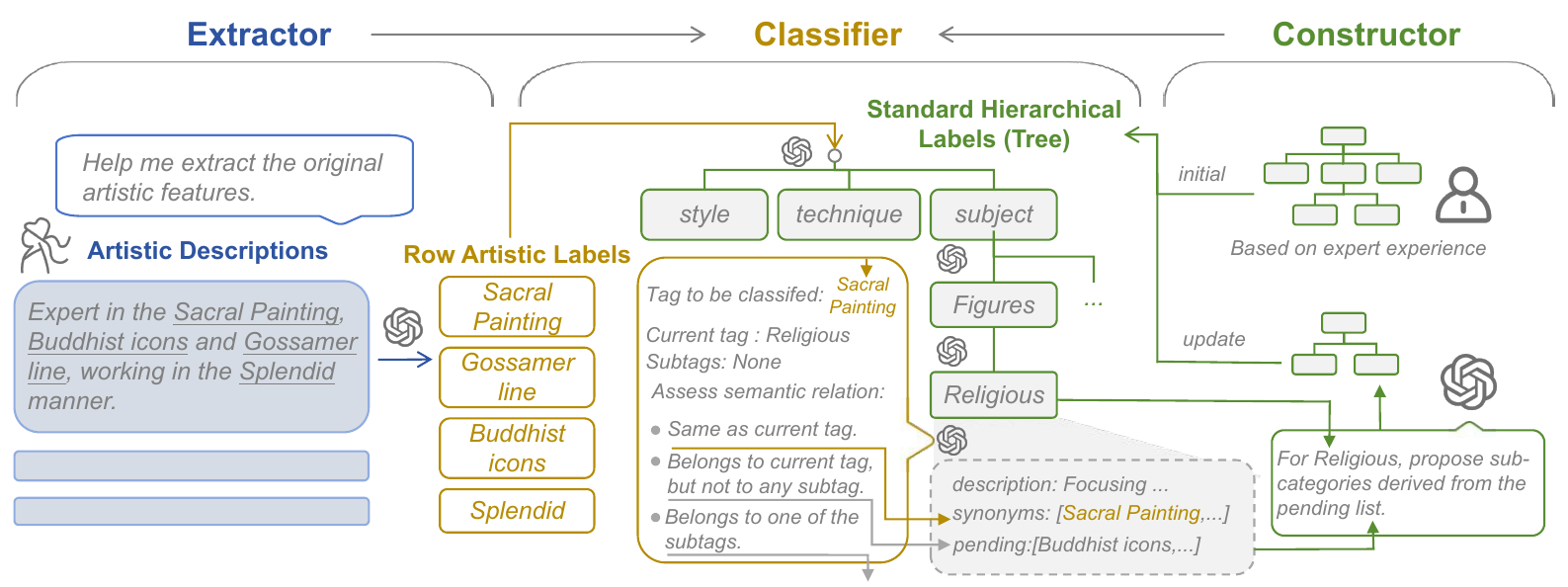}
    \caption{Overview of the multi-scale artistic feature construction pipeline. The LLM serves as Extractor to extract raw semantic labels from unstructured text, Classifier to categorize them into a standard hierarchical label tree, and Constructor to refine the tree structure.
   }
    \label{fig:Hierarchical Labels}
\end{figure*}

\subsubsection{\textbf{Support sustained cohort exploration}}
To address the difficulty of maintaining sustained exploration (C5), the system should provide effective cohort management capabilities to organize and track multiple cohorts and their analytical contexts throughout the research process.

%\hangindent=2.5em 
\textbf{T7: Provide effective cohort management.} The system should support creating, editing, nesting, focusing on, and comparing multiple painter cohorts, while clearly maintaining relationships and analytical contexts among them.

\subsection{Our Workflow}
\label{sec:workflow}
% 基于上述设计需求和任务，我们提出了新的群体分析工作流，如图2右侧所示。我们没有改变研究者的传统研究思路，但是在几个关键的环节改变了研究的方法，帮助研究者提高了研究效率和准确性。

As shown on the right side of Fig.~\ref{fig:Workflow}, our workflow improves several key steps for higher efficiency and accuracy.
%To meet these design requirements and tasks, we introduce our improved cohort analysis workflow, as shown on the right side of Fig.~\ref{fig:Workflow}. We enhance several key steps to improve research efficiency and accuracy.
During the \textit{organize painter features} stage, it converts raw data into structured, multi-scale features. Artistic style features are extracted and organized through LLM-based algorithms (T1), and the entangled inheritance network is reconstructed into a clear structure that reveals lineage relationships (T2).

% In cohort definition and analysis, researchers follow an iterative cycle of ``\textit{formulate cohort hypotheses} $\rightarrow$ \textit{define candidate cohorts} $\rightarrow$ \textit{validate hypotheses}''. The system supports this cycle through two complementary loops. In the visualization loop, multi-dimensional feature distributions help researchers formulate hypotheses and define cohorts; once cohorts are defined, their compositions are visualized across dimensions to support analysis and validation (T3, T4). In the recommendation loop, the system identifies core features of a defined cohort for rapid hypothesis validation and evaluates all painters by their similarity to the cohort. This supports consistency checking, candidate discovery, and further cohort refinement (T5, T6).

In cohort definition and analysis, researchers follow the iterative cycle of ``\textit{formulate cohort hypotheses} $\rightarrow$ \textit{define candidate cohorts} $\rightarrow$ \textit{validate hypotheses}''. The system improves this cycle through two complementary approaches. In the visualization loop, multi-dimensional feature distributions provide feature insights that help researchers formulate hypotheses and define cohorts; once cohorts are defined, their compositions are visualized across dimensions to support analysis and validation (T3, T4).
In the recommendation loop, the system identifies core features of a defined cohort to support rapid hypothesis validation; based on these features, it evaluates the similarity of all painters to the defined cohort. It enables verification of internal consistency and discovery of external candidates, inspiring new hypotheses and cohort refinement (T5, T6).

Throughout the research process, the system provides effective cohort management (T7), allowing users to track multiple cohorts along with their relationships and analytical contexts, thereby supporting sustained cohort exploration.

\section{Methodology}
To support the tasks in Section~\ref{sec:requirement}, we develop three core computational methods: multi-scale artistic feature construction (T1), inheritance relationship reconstruction (T2), and a recommendation model for cohort analysis (T5, T6). These methods provide structured data representations and computational guidance throughout the analytical workflow.

\subsection{Multi-scale Artistic Feature Construction}
\label{Multi-scale Artistic Feature Construction}

% 4.1 构建多层级艺术风格特征
% 我们提出了一套基于大语言模型（LLM）的多层级艺术风格特征构建方法。在该流程中，LLM 在不同阶段承担不同功能角色，分别作为特征提取器（Extractor）、层级分类器（Classifier）以及分类树构建器（Constructor）。整体流程如图 X 所示。
Large language models (LLMs) are employed to construct multi-scale artistic features (T1). As shown in Fig.~\ref{fig:Hierarchical Labels}, the LLM serves 3 functional roles: extractor, classifier, and constructor.

% 提取器（Extractor）. 输入画家的艺术特征描述文本，利用 LLM 从非结构化文本中自动抽取与艺术风格相关的原始语义标签。
\textbf{Extractor.}
Given textual descriptions of a painter's artistic features, the LLM-based Extractor automatically extracts raw semantic labels related to artistic styles from unstructured texts.

% 层级分类器（Classifier）. 在给定标准层级标签树的基础上，采用层级下降策略对原始语义标签进行逐层归类。具体而言，每个原始标签从标签树的根节点出发，若其语义能够匹配某一子节点，则被下推至该子节点，并递归进行分类；最终，每个原始语义标签都会被映射到标签树中的某一具体节点，该节点对应的标准层级标签即作为其对应的艺术风格特征。
\textbf{Classifier.}
Given a standard hierarchical label tree, we adopt a top-down strategy to categorize raw semantic labels. Each label starts from the root node and is recursively pushed to match child nodes based on semantics. Eventually, each raw label is mapped to a specific node in the tree, and the corresponding hierarchical label serves as its standardized artistic style features.
% 标准层级标签树中的每一个标签节点都包含以下三个属性：1）description：记录该标签的语义说明，在分类过程中作为提示词的补充信息，以增强 LLM 对标签语义边界的理解；2）common：用于存储与当前标签在语义上等价的原始语义标签列表，例如“道释画”与”宗教画“；3）uncategorized：用于记录语义上属于当前标签的子概念，但尚未被现有子节点覆盖的原始标签。在分类过程中，原始标签将被记录在对应的节点的common或uncategorized属性中。
Each node in the standard hierarchical label tree contains three attributes: 
(1) \textit{description}, which provides semantic explanations of the label and serves as prompt context to enhance the LLM's understanding of semantic boundaries; 
(2) \textit{synonyms}, which stores semantically equivalent raw labels (e.g., ``Taoist-Buddhist painting'' and ``religious painting''); 
and (3) \textit{pending}, which records raw labels that are subconcepts of the current node but are not yet covered by existing child nodes. During classification, each raw label is stored in either the \textit{synonyms} or \textit{pending} field of the corresponding node.

\begin{figure*}[!t]
    \setlength{\belowcaptionskip}{-0.3cm}
    \centering 
    \includegraphics[width=0.95\textwidth]{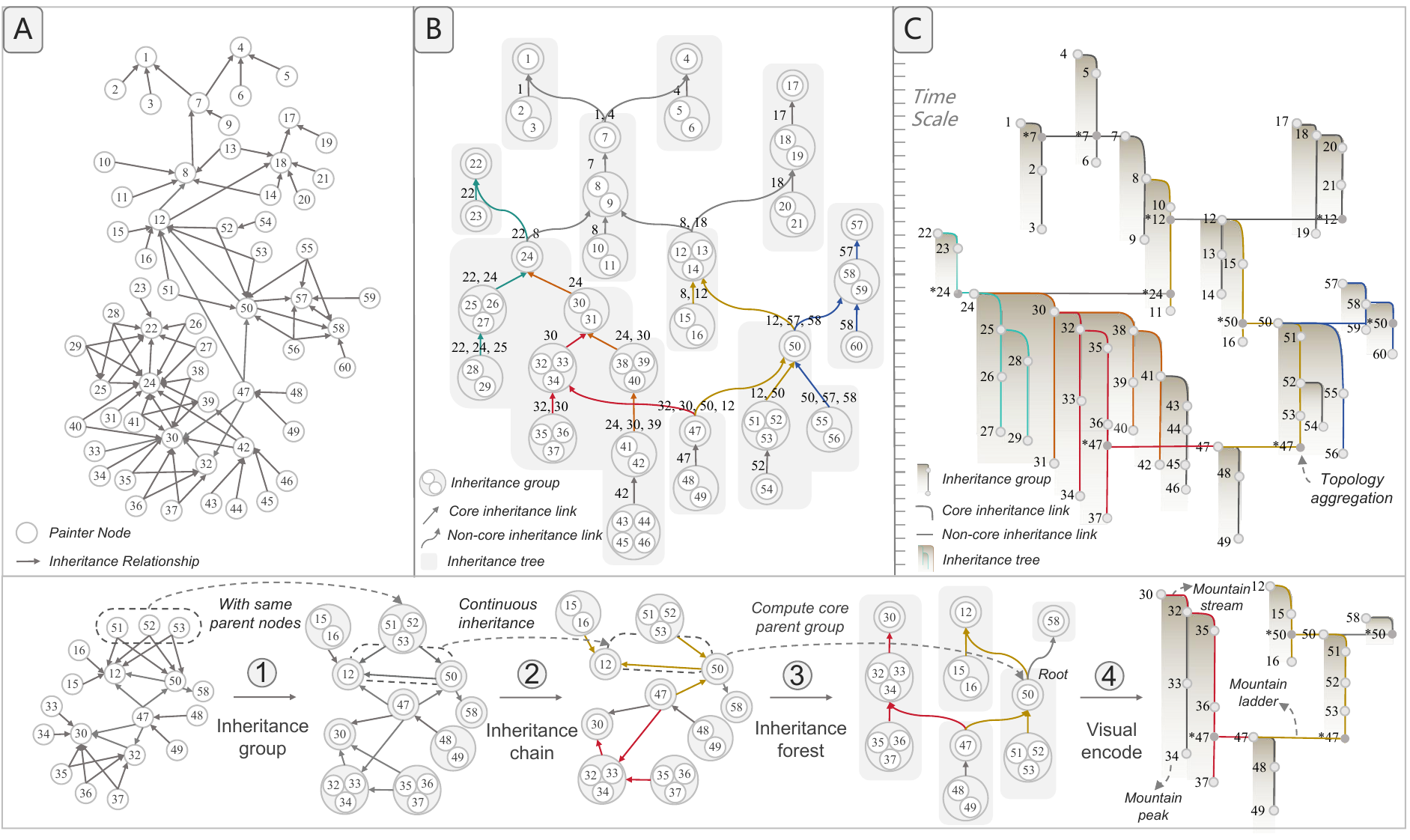}
    \caption{Overview of the inheritance relationship reconstruction algorithm. The top row shows the full pipeline applied to a 60-node subset from our dataset: (A) the original node-link graph of inheritance relationships; (B) the reconstructed forest structure, where large circles represent inheritance groups, colored paths denote inheritance chains (black for single-parent chains), straight lines indicate core inheritance links, curved lines indicate non-core inheritance links, and light gray backgrounds delineate inheritance trees; (C) the final Inheriting Mountain View encoding. The bottom row illustrates the reconstruction steps in detail using a smaller subset: \ding{192} grouping painters with shared predecessors into inheritance groups; \ding{193} merging multi-source relationships into continuous inheritance chains; \ding{194} identifying core parent groups to construct the inheritance forest; and \ding{195} visual encoding into the Inheriting Mountain View.}
    \label{fig:Relation Reconstruction}
\end{figure*}

% 标准层级标签树构建器（Constructor）. 支持用户指定与LLM推荐两种方式构建标签结构。用户指定方式可约束标签体系的范围，从而避免偏离研究目标；在本文中，标准层级标签树的高层结构由领域专家定义。
\textbf{Constructor.} The standard hierarchical label tree can be built through user specification or LLM-based recommendation. User specification helps constrain the label space and prevent deviation from research objectives; in this study, the high-level structure of the label tree is defined by domain experts.

% 对于需要进一步细化的标签分支，当用户难以依据经验明确子类别划分时，引入 LLM 作为辅助，根据父标签及其 uncategorized 中包含的原始语义标签，自动推荐合适的下级层级标签。加入子标签后，uncategorized 中的原始标签将再次通过层级分类器进行下推与归类。通过上述过程，分类树结构能够根据数据分布与分析需求进行演进，从而不断完善其层级结构。

For label nodes needing refinement, users may not know subcategories in advance. For example, under a high-level subject category such as ``religious painting,'' the specific sub-subjects (e.g., Buddhist figures, ghosts and deities) depend on what actually appears in the data. In such cases, the LLM recommends appropriate child labels based on the parent label and the raw labels in its \textit{pending} attribute. Once new child labels are added, the raw labels in \textit{pending} are reprocessed by the Classifier for further categorization. Through this iterative process, the label tree progressively refines its hierarchical structure according to data distribution and analysis needs.

Through the above three components, each painter's unstructured textual descriptions are transformed into standardized multi-scale artistic style labels in a hierarchical label tree. Detailed prompt templates, examples, and validation details are provided in the supplementary material.

% 对于某一画家，在获得其多层级标准标签后，我们进一步计算各标签相对于该画家的重要性。具体而言，我们综合考虑语义贡献度与累积频次两个因素来衡量标签的重要性。首先，借鉴 [X] 的方法，对标准标签对应的原始标签所在文本进行掩码（mask）操作，通过比较掩码前后文本语义表示的变化来量化该标签的语义贡献度；随后，将同一画家中映射到同一层级标签的贡献度进行累加，并在所有标签之间进行归一化，得到最终的重要性分值。基于该分值，可通过设定阈值（默认 0.7）筛选出该画家的代表性艺术风格特征。
We compute the importance of each feature label for all painters. Since the feature label may correspond to multiple raw labels, we first quantify the semantic contribution of each raw label by masking it in its source text and measuring the resulting change in semantic representation, inspired by ~\cite{zhang2022mderank}. We then accumulate the contributions of all raw labels mapped to the same feature label, and normalize the scores among all these labels. Labels exceeding a threshold (default 0.7) are selected as the painter's representative artistic style features.

% 4.2 传承关系重构 Inheritance relationship reconstruction
\subsection{Inheritance relationship reconstruction}
\label{Inheritance relationship reconstruction}

% 我们提出一种师承关系重构算法，将错综复杂的多父继承有向无环图转换为清晰的森林结构。具体而言，首先基于画家间相同的师承来源构建继承组，对具有一致继承模式的个体进行统一表示；随后，将多源继承关系组织为连续的继承链，以刻画传承路径；在此基础上，通过识别核心继承脉络，构建具有层级结构的继承森林，从而为后续可视分析提供结构化支撑。
We propose an inheritance relationship reconstruction algorithm (T2) that transforms a complex multi-parent inheritance DAG into a clear forest (Fig.~\ref{fig:Relation Reconstruction}). Specifically, we first construct \textit{inheritance groups} based on shared inheritance sources among painters, providing a unified representation for individuals with consistent inheritance patterns. Next, multi-source inheritance relationships are organized into continuous \textit{inheritance chains} to characterize transmission paths. Finally, we identify the core inheritance lineage for each group and construct a hierarchical \textit{inheritance forest}, enabling researchers to identify lineage groupings, trace dominant inheritance paths, and understand relationships among different lineages.

% 4.2.1 继承组 Inheritance Group
\subsubsection{\textbf{Inheritance Group}}

Given an original inheritance network \(G=(V,E)\), where each node \(v \in V\) represents a painter and each directed edge \((v_i, v_j) \in E\) indicates that painter \(v_i\) inherited from a predecessor \(v_j\). For any painter node \(v_i\), the set of its predecessors is defined as:

\[
P(v_i) = \{ v_j \in V \mid (v_i, v_j) \in E \}.
\]

% 在师承关系网络中，若一组画家具有相同的师傅集合Si，则它们在传承结构上具有同质性，将这些节点聚合在一起，能够减少连接，同时形成群体的隐喻。定义一个继承组为
If a group of painters share the same predecessor set $S_i$, they exhibit structural homogeneity in the inheritance network. Aggregating these nodes reduces redundant connections while forming a natural grouping (Fig.~\ref{fig:Relation Reconstruction}\ding{192}). An \textit{inheritance group} $g_i$ thus can be defined as:

\[
g_i = \{ v_k \in V \mid P(v_k) = S_i \}.
\]

% 若没有师傅，则该节点独立构成一个继承组。
% 继承组将作为传承结构中的基本单元，统一表达具有相同继承来源的画家个体。

If a painter has no predecessors ($P(v) = \emptyset$), it forms a singleton group. Inheritance groups serve as the fundamental units in the reconstructed structure, unifying painters with identical predecessor sets.

% 4.2.2 继承链 Inheritance Chain
\subsubsection{\textbf{Inheritance Chain}}

An inheritance group may inherit from multiple parent groups, and inheritance relationships may also exist among these parent groups. For example, a group \(g\) may inherit from parent groups \(A\) and \(B\), while \(B\) itself inherits from \(A\). In the context of art history, such a structure is typically interpreted as a continuous inheritance path \(A \rightarrow B\), rather than two independent sources.

% 因此，我们将父组集合中的传承关系组织为若干条按照传承方向自底向上排列的序列，并定义其为继承链（Inheritance Chain）。对于传承组 g，其父组集合记为 P(g)，对应的继承链集合可表示为：

Therefore, we re-organize the inheritance relationships among parent groups into ordered sequences along the inheritance direction. We denote this ordered sequence as an \textit{inheritance chain} (Fig.~\ref{fig:Relation Reconstruction}\ding{193}), which is represented as: 
\[
ch_i = (g_1 \rightarrow g_2 \rightarrow \cdots \rightarrow g_m), \quad g_j \in P(g).
\]

\noindent where \(P(g)\) denotes the set of all parent groups of the inheritance group \(g\). Within a chain, each group inherits from all upstream groups. Inheritance chains serve as connective relationships in the reconstructed structure, representing inheritance paths among groups.

Finally, the corresponding set of inheritance chains is:
\[
Ch(g) = \{ ch_1, ch_2, \ldots, ch_k \}.
\]

% 4.2.3 继承森林 Inheritance Forest
% 为构建具有层级结构的继承森林，需要在继承链的基础上为每个继承组识别其核心传承来源。

\subsubsection{\textbf{Inheritance Forest}}
\label{Inheritance Forest}

To construct the inheritance forest, we need to identify a core parent group for each inheritance group based on its inheritance chains.
% 对于继承组 g，其父组集合记为 P(g)。给定父组 p，定义其对 g 的覆盖度为沿继承链方向从 p 出发可达的父组数量。
For an inheritance group $g$, we first evaluate how well each parent group $g_j \in P(g)$ represents the overall inheritance of $g$. We define $N(g_j, g)$ as the number of groups in $P(g)$ that are reachable from $g_j$ along the inheritance chains, including $g_j$ itself, and define the coverage ratio as $R(g_j, g) = N(g_j, g) / |P(g)|$. Given a threshold $\tau$ (default $0.65$), the core parent group $g^*$ of $g$ is:

\[
g^* = 
\begin{cases}
\arg\max_{g_j \in P(g)} R(g_j, g), & \text{if } \max_{g_j} R(g_j, g) > \tau, \\[6pt]
\emptyset, & \text{otherwise.}
\end{cases}
\]

When $g^* = \emptyset$, $g$ is treated as a new root group. By determining a unique core parent group for each inheritance group, the original inheritance graph is reconstructed into an \textit{inheritance forest} composed of multiple hierarchical trees (Fig.~\ref{fig:Relation Reconstruction}\ding{194}), where inheritance groups serve as structural units and inheritance chains characterize group connections.

\begin{figure*}[!t]
  \centering
  \includegraphics[width=0.95\textwidth]{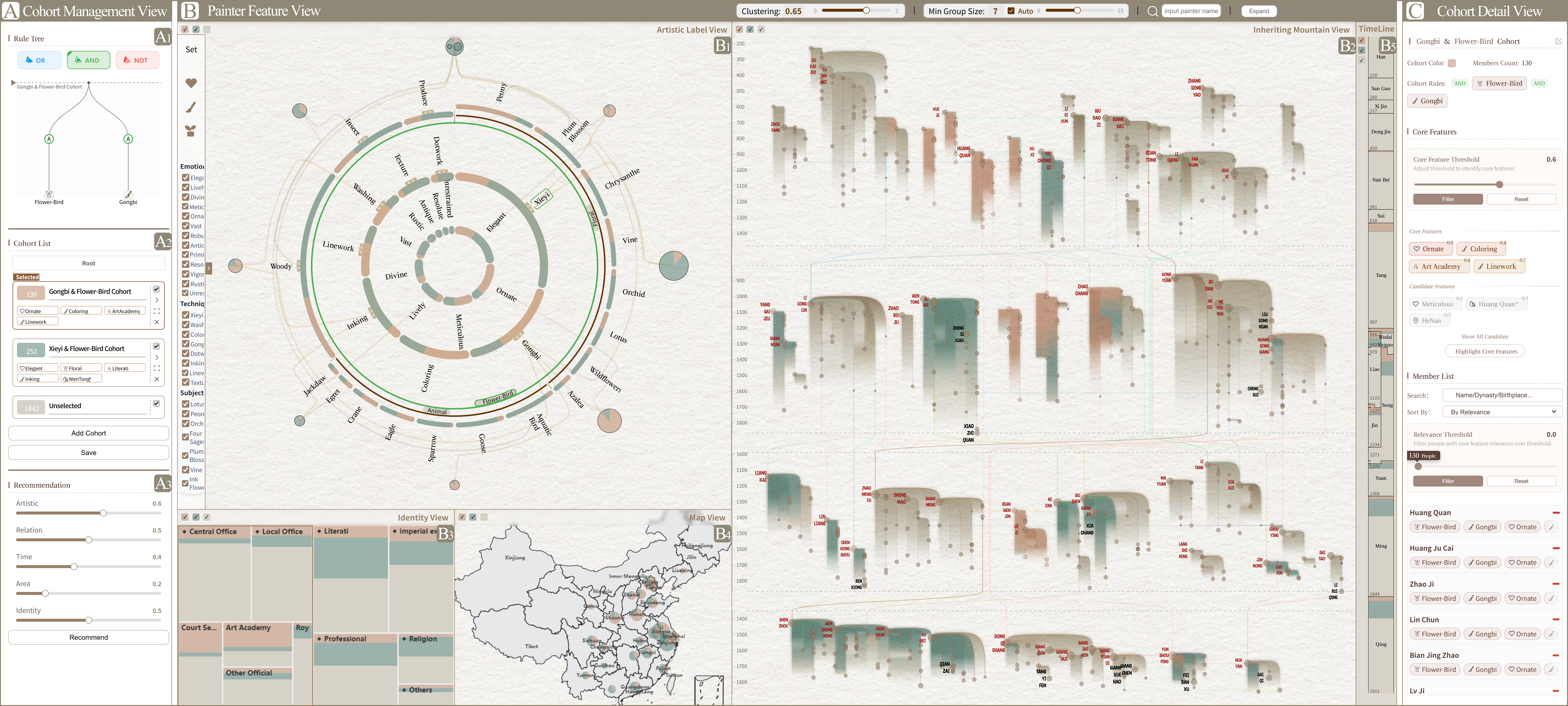}
  \caption{The interface of {\name} for interactive exploration and analysis of Chinese historical painter cohorts. It consists of three main views: (A) Cohort Management View, including the Rule Tree (A1), Cohort List (A2), and Recommendation (A3) panels for cohort definition and management; (B) Painter Feature View, visualizing multi-dimensional feature distributions through five coordinated subviews: Artistic Label View (B1), a foldable doughnut chart with inner thematic rings and outer circle packs to reveal hierarchical artistic style labels and their co-occurrence patterns; Inheriting Mountain View (B2), a landscape-metaphor visualization encoding inheritance groups and inheritance links to reveal lineage structures; Identity View (B3), Map View (B4), and Timeline View (B5), showing the social identity, geographic, and temporal distributions of painters respectively; and (C) Detail View, providing detailed information about selected cohorts and features. A dynamic explanation of our interface is provided in the supplementary demo video. }
\label{fig:System}
\end{figure*}

\subsection{Recommendation Model} % 推荐模型
\label{Recommendation Model}
% Inspired by prior work on traditional Chinese painting school analysis that characterizes painters through art-theoretical features~\cite{sijia2024tcpvis}, we formulate cohort recommendation as a feature-relevance estimation problem. 
Since a cohort may involve a large feature space, we first identify core features of a cohort, and use them to compute the relevance with painters. This relevance is then employed to not only recommend external potential painters but also evaluate the feature consistency of existing cohort members (T5, T6).
% The recommendation model supports two core tasks: identifying the core features of a defined cohort and assessing the relevance of painters to the cohort.

\subsubsection{\textbf{Core Feature Identification for Cohorts}}
\label{core-feature-identification-for-cohorts}

Each painter $p$ possesses a set of hierarchical features across multiple dimensions, including artistic style, inheritance, identity, geography, and time, denoted as $T(p) = \{t_1, t_2, \dots, t_n\}$. Each feature $t = (l_1, l_2, \dots, l_h)$ is represented as a label path from high to low levels, where $l_i$ is the label at level $i$, and $h$ denotes the depth of the feature. For a painter cohort $C$, let $T_C$ denote the set of all features within the cohort. 

To quantify the representativeness of each feature $t$ for the cohort, we consider its frequency, dimensional distribution, and hierarchical depth, to define its Representativeness Score (RS):

\[
RS(t) = f(t) \cdot CV_d \cdot w_h(t),
\]
where
%\begin{itemize}
$f(t)$ is the occurrence frequency of feature $t$ in the cohort, reflecting its prevalence:
    \[
    f(t) = \sum_{p \in C} \sum_{t' \in T(p)} \mathbb{I}(t \preceq t'),
    \]
% \noindent where $\mathbb{I}(\cdot)$ is the indicator function and $t \preceq t'$ means $t$ is a prefix of $t'$, so that each specific feature also counts toward its corresponding higher-level features, enabling multi-scale statistics.
\noindent where $\mathbb{I}(\cdot)$ is the indicator function and $t \preceq t'$ means $t$ is a prefix of $t'$, so each specific feature also contributes to its higher-level features, enabling multi-scale analysis.
    
$CV_d$ is the coefficient of variation for dimension $d$, measuring feature dispersion in that dimension. A higher value indicates greater differences in feature frequencies within that dimension, implying higher discriminability and greater contribution to core feature identification. $w_h(t)$ is a hierarchical weight capturing the semantic specificity of the feature:
% $CV_d$ is the coefficient of variation for dimension $d$, measuring feature dispersion in that dimension. Higher values indicate larger frequency differences and thus greater discriminability, leading to stronger contributions to core feature identification. $w_h(t)$ is a hierarchical weight capturing the feature's semantic specificity:
    \[
    w_h(t) = \frac{\text{depth}(t)}{H_{\max}},
    \]
\noindent where $\text{depth}(t)$ is the depth of feature $t$ (root depth $=1$), and $H_{\max}$ is the maximum depth of the feature hierarchy. Deeper features are more specific and thus weighted more highly.
% \noindent  where $\text{depth}(t)$ is the depth of feature $t$ (root depth $= 1$), and $H_{\max}$ is the maximum depth of the feature hierarchy. Deeper features represent more specific semantics and receive higher weights.
%\end{itemize}

After computing representativeness scores for all features, we exclude features that are guaranteed by the cohort definition and shared by all cohort members (e.g., features introduced by AND rules), as they do not provide additional discriminative information for characterizing the cohort. Among the remaining features, those whose scores exceed a default threshold (default 0.7) are selected as the core feature set $F_C$ of the cohort $C$. Users can adjust this threshold to control the number of recommended core features.
% After computing representativeness scores for all remaining features, those exceeding a default threshold (default 0.7) are selected as the core feature set $F_C$ of the cohort $C$. Users can adjust this threshold to control the number of recommended core features.
% Note that features that are shared by all cohort members (e.g., features added as AND rules in the cohort definition) will be selected as core features, as they do not provide discriminative information for characterizing the cohort.
% After computing representativeness scores for all features, they are ranked in descending order, and the top $K$ features (default $K=6$) are selected as the core feature set $F_C$ of the cohort $C$.

\subsubsection{\textbf{Relevance-based Painter Recommendation}}
\label{Relevance-based Painter Recommendation}

A painter's relevance to cohort $C$ is assessed by how well its features match the cohort's core features $F_C$:

% The relevance of a painter $p$ to a cohort $C$ can be assessed by measuring how well the painter's features match the cohort's core features $F_C$:
\[
\text{Rel}(p, C) = \sum_{d \in D} \lambda_d \sum_{t \in F_C(d)} RS(t) \cdot \text{Match}(t, p),
\]

% where $F_C(d)$ denotes core features in dimension $d \in D$, and $\lambda_d$ is the dimension weight, defaulting to the normalized coefficient of variation to emphasize highly discriminative dimensions. Users can manually adjust $\lambda_d$ for different analytical objectives. $\text{Match}(t, p)$ measures how well painter $p$ matches core feature $t$:

where $F_C(d)$ denotes the subset of core features in feature dimension $d \in D$, and $\lambda_d$ is the dimension weight, defaulting to the normalized coefficient of variation to emphasize highly discriminative dimensions. Users can manually adjust $\lambda_d$ to achieve different analytical objectives. $\text{Match}(t, p)$ measures how well painter $p$ matches core feature $t$:
    \[
    \text{Match}(t, p) = \max_{t' \in T(p)} \frac{\text{LCP}(t, t')}{\text{depth}(t)},
    \]
\noindent where $T(p)$ is the feature set of painter $p$ and $\text{LCP}(t, t')$ is the length of the longest common prefix between $t$ and $t'$. This ensures that painters matching core features at deeper levels receive higher relevance scores.

For a cohort $C$, $\text{Rel}(p, C)$ is computed for all painters. External painters whose relevance exceeds a default threshold (default 0.6) are recommended as potential candidates for the cohort. Users can also adjust this threshold interactively to control the scope of recommendation. The relevance of internal painters is used to evaluate the feature consistency of the cohort.

\section{Visual Analytics System}

\begin{figure*}[!t]
    \setlength{\belowcaptionskip}{-0.3cm}
    \centering 
    \includegraphics[width=0.95\textwidth]{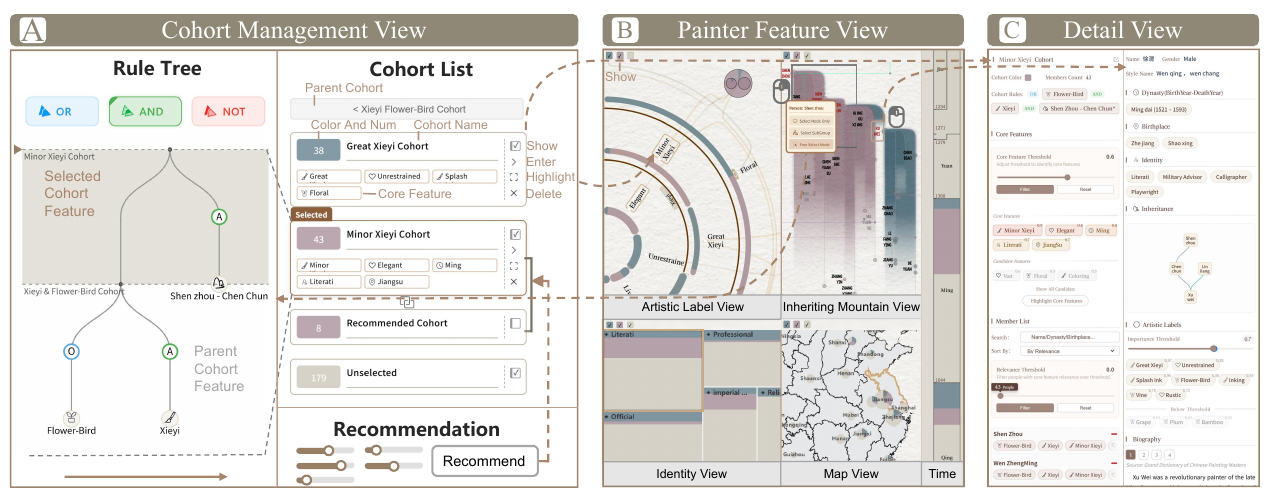}
    \caption{Interaction design of HPC-Vis. The Cohort Management View (A) consists of the Rule Tree, Cohort List, and Recommendation panel. The Rule Tree combines feature rules with selectable logical operators (OR, AND, NOT) to define the selected cohort, with parent cohort rules separated at the bottom. The Cohort List displays cohort cards with key information; the selected cohort's details are shown in the Detail View (C). The Recommendation panel generates a recommended cohort for the selected cohort. The Painter Feature View (B) presents multi-dimensional feature distributions through five coordinated subviews. Cohort colors from the Cohort List are reflected within each feature element, with color proportions indicating the share of different cohorts. Right-clicking features adds rules to the Rule Tree; left-clicking inspects details in the Detail View (C).
   }
    \label{fig:System Interaction}
\end{figure*}
We propose \textbf{HPC-Vis}, a visual analytics system supporting the improved workflow in Section~\ref{sec:workflow} for interactive painter cohort exploration. As shown in Fig.~\ref{fig:System}, it consists of three views: \textit{Cohort Management View} (A) for cohort management, \textit{Painter Feature View} (B) for multi-dimensional feature distributions, and \textit{Detail View} (C) for feature and cohort details.

% We propose \textbf{HPC-Vis}, a visual analytics system that supports the improved workflow described in Section~4.2 for interactive exploration of painter cohorts. Fig.~\ref{fig:System} shows that it consists of three views: \textit{Cohort Management View} (A) for organizing and managing cohorts, \textit{Painter Feature View} (B) for visualizing multi-dimensional feature distributions, and \textit{Detail View} (C) for inspecting detailed information of features and cohorts.

\subsection{Cohort Management View}
The Cohort Management View (Fig.~\ref{fig:System}A) provides unified cohort organization and management (T7), consisting of three components: the \textit{Rule Tree}, the \textit{Cohort List}, and the \textit{Recommendation} panel. Fig.~\ref{fig:System Interaction} illustrates the interactions among these components and other views.

\textbf{Rule Tree} visualizes the feature-based painter filtering rules for cohort initialization. It is organized as a tree structure where feature labels serve as leaf nodes and logical operators serve as internal nodes. Painters satisfying these rules are picked out for the current cohort. Three logical operators are supported: OR, AND, and NOT (colored blue, green, and red, respectively). Users can switch the active operator via buttons at the top; once selected, the mouse cursor changes accordingly to indicate the current mode. Users can right-click features in the Painter Feature View to add or remove rules. Rules are arranged from left to right in the adding order. Rules inherited from the parent cohort are displayed at the bottom and separated from the selected cohort's newly added rules by a dashed line.
%, helping users maintain the analytical context between parent and child cohorts.

\textbf{Cohort List} supports hierarchical navigation among parent and child cohorts. The top navigation button returns users to the parent cohort, while the list shows all direct child cohorts. The selected cohort is highlighted with a brown border. Each cohort card shows its color, painter count, name, and core features; the core features are recommended by default (Section~\ref{core-feature-identification-for-cohorts}) and can be manually adjusted. Card buttons allow users to toggle visibility, highlight core features in the Painter Feature View, navigate to sub-cohorts, or delete the cohort. Painters not assigned to any listed cohort automatically form a special unassigned group with a light brown color, which dynamically updates as cohorts change. Users can inspect each cohort's members in the Detail View.

% \textbf{Cohort List} supports hierarchical navigation among parent and child cohorts. At the top of the list, a navigation button labeled with the parent cohort's name allows users to return to the upper level. The list area shows all direct child cohorts under the  parent cohort. The selected cohort is highlighted with a brown border, and users can click to switch the selection. Each cohort is presented as a card: the upper-left corner shows the cohort color and the number of painters; the top displays the cohort name; and the middle section shows the cohort's core features, which are recommended as described in Section~5.3.1 by default and can be manually adjusted. Buttons on the right side of each card allow users to toggle the cohort's visibility, highlight the cohort's core features with brown outlines in the Painter Feature View, navigate to its sub-cohorts, or delete the cohort. Painters not assigned to any listed cohort automatically form a special unassigned group with a light brown color, which dynamically updates as cohorts change. Users can inspect the member list of each cohort in the Detail View.

\textbf{Recommendation Panel} recommends painters for the selected cohort based on the relevance scores computed in Section~\ref{Relevance-based Painter Recommendation}, with default dimension weights derived from feature concentration. Weight sliders let users steer recommendations across feature dimensions. After recommendation, the system adds a same-color recommended cohort to the Cohort List; users can examine its distributions, inspect recommended painters in the Detail View, and merge them as needed.

% \textbf{Recommendation Panel} recommends more painters for the selected cohort based on the relevance scores computed in Section~5.3.2. It contains weight sliders for multiple feature dimensions. Users can manually adjust these weights to control the recommendation direction. Upon clicking the recommendation button, the system generates a recommended cohort in the Cohort List, using the same color as the selected cohort. Users can toggle its visibility in the Painter Feature View to examine its feature distributions across views for verification, and inspect the recommended painters in the Detail View and merge them into the selected cohort as needed.

% 5.3 Painter Feature View
\subsection{Painter Feature View}
The Painter Feature View (Fig.~\ref{fig:System}B) visualizes multi-dimensional feature distributions (T3, T4) through five subviews: \textit{Artistic Label View}, \textit{Inheriting Mountain View}, \textit{Identity View}, \textit{Map View}, and \textit{Timeline}.
As shown in Fig.~\ref{fig:System Interaction}, the Painter Feature View is coordinated with the Cohort List. It displays the multi-dimensional feature distributions of all painters within the parent cohort that users are currently navigating. Each feature is an area-based element whose color proportions show visible cohort shares, enabling intuitive comparison. In each subview, users can left-click any feature to inspect its details in the Detail View, and right-click to add it as a rule for the selected cohort. Mouse wheel scrolling zooms the view; holding the middle mouse button and dragging up or down expands or collapses hierarchical features.
% Each feature is represented as an area-based element, in which the color proportions reflect the share of each visible cohort, enabling intuitive comparison of cohort features.

% 5.3.1 Artistic Label View
\begin{figure*}[!t]
    \setlength{\belowcaptionskip}{-0.3cm}
    \centering 
    \includegraphics[width=0.95\textwidth]{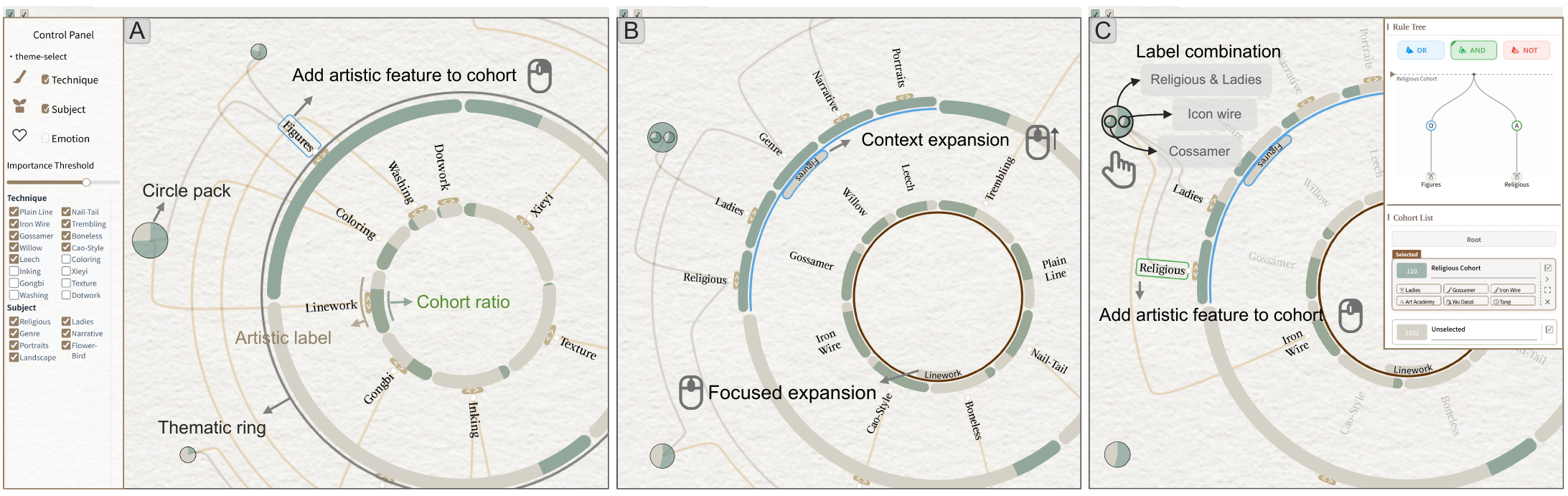}
    \caption{
    The Artistic Label View, illustrated through an exploration of figure painters. 
    % The view consists of inner thematic rings showing artistic feature distributions and outer circle packs revealing label co-occurrence patterns. A left control panel allows customization of visible themes and label importance thresholds. 
    (A) The ``Figures'' label is added to the cohort via right-click, and the resulting cohort (green) shows a notable proportion in the ``Linework'' technique. (B) To explore finer granularity, ``Figures'' is expanded via Context expansion, and ``Linework'' is expanded via Focused expansion, revealing detailed sub-label distributions. (C) The ``Religious'' sub-label is further added with an AND rule, refining the cohort to religious figure painters. The circle pack reveals a prominent combination ``Religious \& Ladies'' containing two nested sub-combinations associated with ``Iron Wire'' (a vigorous, angular line technique) and ``Gossamer'' (a delicate, flowing line technique) respectively, suggesting two distinct technical tendencies within the cohort worth further exploration.
   }
    \label{fig:Artistic Label View}
\end{figure*}

\textbf{1) Artistic Label View}

We design a novel \textit{foldable doughnut chart} to visualize hierarchical artistic style labels (Fig.~\ref{fig:System}B1). It combines inner \textit{thematic rings} for label distributions and outer \textit{circle packs} for label co-occurrences (Fig.~\ref{fig:Artistic Label View}).

% Each thematic ring corresponds to one of the three major artistic themes in Chinese painting (subject, technique, and emotion), which also serve as the top-level categories in the hierarchical label tree (Section~\ref{Multi-scale Artistic Feature Construction}). Each arc within a ring represents an artistic label, with arc length encoding the proportion of painters possessing that label within the corresponding theme. For hierarchical labels, the system provides two expansion modes (Fig.~\ref{fig:Artistic Label View}B). In \textit{Context expansion}, holding the middle mouse button and dragging upward expands a label: its arc contracts into a thin inner-labeled line, while child labels appear outside for cross-level comparison. In \textit{Focus expansion}, middle-clicking a label restricts the thematic ring to its child labels for focused inspection. Dragging downward collapses expanded labels.

Each thematic ring corresponds to one of the three artistic themes in Chinese painting (subject, technique, and emotion), which also serve as the top-level categories in the hierarchical label tree (Section~\ref{Multi-scale Artistic Feature Construction}). Each arc within a ring represents an artistic label, with arc length encoding the proportion of painters possessing that label within the theme. For hierarchical labels, the system provides two expansion modes (Fig.~\ref{fig:Artistic Label View}B). In \textit{Context expansion}, holding the middle mouse button and dragging upward expands a label: its arc contracts into a thin inner-labeled line, while child labels appear outside for cross-level comparison. In \textit{Focus expansion}, middle-clicking a label restricts the thematic ring to its child labels for focused inspection. Dragging downward collapses expanded labels.

% The first is \textit{Context expansion}: holding the middle mouse button and dragging upward expands a label. The label arc contracts into a thin arc line with the label text on its inner side, while child labels are displayed outside, facilitating comparison across hierarchical levels. The second is \textit{Focus expansion}: middle-clicking a label restricts the corresponding thematic ring to show only that label's child labels, enabling focused inspection of the sub-label distribution. Collapsing is performed by holding the middle mouse button and dragging downward.

Each circle in the circle packs represents a significant label combination, automatically extracted through closed frequent itemset mining~\cite{zaki2002charm}. Circles with a more specific label combination are nested inside the circle of their broader combination. For example, the combination ``Religious \& Ladies'' contains nested circles such as ``Religious \& Ladies \& Iron Wire'' and ``Religious \& Ladies \& Gossamer'' (Fig.~\ref{fig:Artistic Label View}C).

% Each circle in the circle packs represents a significant label combination, automatically extracted through closed frequent itemset mining~\cite{zaki2002charm}. Circle size encodes the number of painters with that combination. Subset relationships are shown as nested circles, with more specific combinations placed inside broader ones. For example, ``Religious \& Ladies'' contains nested circles such as ``Religious \& Ladies \& Iron Wire'' and ``Religious \& Ladies \& Gossamer'' (Fig.~\ref{fig:Artistic Label View}C).

Users can right-click a label or a combination circle to add it as a rule for the selected cohort (Fig.~\ref{fig:Artistic Label View}A). For a single label, the rule requires ``having this label''; for a combination, it requires ``having all labels in this combination.'' Selected items are highlighted with outlines colored by the active logical operator. Colors in label and combination circles reflect the proportion of painters from different cohorts.

% Curved links connect each combination with its constituent labels; hovering over either side highlights the related labels or combinations (Fig.~\ref{fig:Artistic Label View}C).

Curved links connect each combination with its constituent labels. Bidirectional highlighting is supported (Fig.~\ref{fig:Artistic Label View}C): hovering over a label or combination highlights all related combinations or labels.

A control panel on the left side allows users to adjust the visibility of themes and labels. It also provides a label importance threshold slider: a painter is considered to possess a label only when its importance score (Section~\ref{Multi-scale Artistic Feature Construction}) exceeds the threshold, enabling users to customize the analytical precision.

\textbf{2) Inheriting Mountain View}

\begin{figure}[!t]
    \setlength{\belowcaptionskip}{-0.3cm}
    \centering 
    \includegraphics[width=0.95\columnwidth]{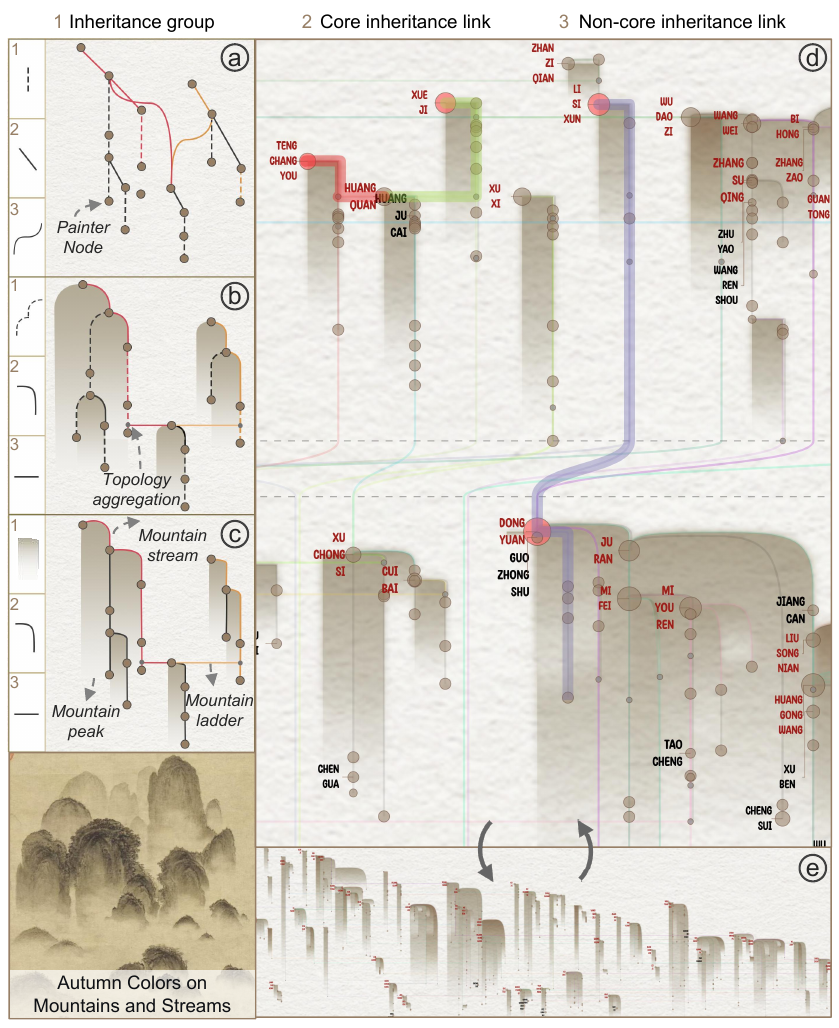}
    \caption{
Design iterations and layout of the Inheriting Mountain View. (a)--(c) Three design iterations for the visual encoding of inheritance 
groups, core inheritance links, and non-core inheritance links. The landscape metaphor is inspired by traditional Chinese mountain paintings (e.g., \textit{Autumn Colors on Mountains and Streams}). (d) Multi-row layout. (e) Single-row layout. 
% Users can switch between the two layouts.
}
    \label{fig:Inheriting Mountain View}
\end{figure}

We design the \textit{Inheriting Mountain View} (Fig.~\ref{fig:System}B2) to represent painter inheritance relationships. Inspired by traditional Chinese landscape paintings, it encodes the inheritance groups, inheritance chains, and inheritance trees reconstructed in Section~\ref{Inheritance relationship reconstruction} into a landscape metaphor, turning complex inheritance networks into an aesthetically meaningful yet readable representation. To achieve this design target, we iteratively explored multiple prototypes, as shown in Fig.~\ref{fig:Inheriting Mountain View}.

The design encodes three elements: (1) \textit{inheritance groups} as the basic structural units, (2) \textit{core inheritance link}, which link each group to its core parent group and form the backbone of the inheritance tree, and (3) \textit{non-core inheritance link}, which represent additional inheritance paths and need to be visually distinguished to avoid obscuring the main tree structure. Each design iteration refines the encoding of these three elements.

% The design addresses three core visual encoding elements: (1) \textit{inheritance groups} as the basic structural units, (2) \textit{core inheritance link}, which link each group to its core parent group and form the backbone of the inheritance tree, and (3) \textit{non-core inheritance link}, which represent additional inheritance paths and need to be visually distinguished to avoid obscuring the main tree structure. Each design iteration refines the encoding of these three elements.

\textbf{Design Iteration 1} (Fig.~\ref{fig:Inheriting Mountain View}a).
Inheritance groups are represented as vertical dashed lines, with painter nodes positioned by historical period. Each inheritance chain starts from the top node of a group and connects to the nearest parent node in its parent group. These inheritance chains are color encoded: chains containing only a single group are shown in black, while multi-group chains are shown in color; chains sharing the same prefix use the same color. Core and non-core inheritance links are distinguished by straight and curved connections, respectively. However, the overall tree structure remains difficult to perceive, and as the number of groups increases, intersecting links produce severe visual clutter.
% Inheritance groups are vertical dashed lines with painter nodes positioned by historical period. Core inheritance links are straight lines, non-core links are curved, and chains sharing a prefix use consistent colors.
% However, the overall tree structure is hard to perceive,and as the number of groups increases, intersecting links cause severe visual clutter.
% Inheritance groups are represented as vertical dashed lines, with painters shown as nodes along the line positioned by historical period. core inheritance links are drawn as straight lines, while non-core inheritance links use curved lines. Chains sharing the same prefix are encoded in consistent colors.

% However, this design has two major limitations: the overall tree structure is difficult to perceive, and as the number of groups increases, intersecting links cause severe visual clutter.

\textbf{Design Iteration 2} (Fig.~\ref{fig:Inheriting Mountain View}b). We introduced a landscape metaphor inspired by traditional Chinese mountain paintings. Inheritance groups are encoded as curved dashed mountain contours rather than simple vertical lines. The layout still follows the vertical organization of groups, but when a node in a group connects to a child group, that node becomes a peak at which the contour branches into two directions: one branch extends toward the child group, and the other continues to represent the remaining members of the current group. In this way, connected groups within the same inheritance tree are visually integrated into overlapping mountain ranges. Core inheritance links are drawn as solid lines embedded in these branching contours. For non-core inheritance links, we introduce topological aggregation: we identify intermediate groups whose predecessor sets match the remaining chain prefix and insert virtual relay nodes so that links with the same prefix merge before diverging. These aggregated links are rendered as horizontal lines resembling mountain ladders, reducing link complexity and making inheritance trees more visually coherent. However, the curved contours blur boundaries between groups within the same tree and make the tree structure unclear.

% To address these issues, we introduce a landscape metaphor inspired by Chinese mountain paintings. Inheritance groups are encoded as segments of continuous mountain outline, where groups within the same tree are concatenated to form layered mountain ranges. When a painter node has a connected child group, it serves as a peak point with the contour branching in two directions, creating a curved mountain profile. Core inheritance links are drawn as curved lines resembling mountain streams. For non-core inheritance links, we identify intermediate groups whose predecessor sets match the remaining chain prefix and insert virtual nodes within them as relay points, allowing chains with identical prefixes to converge. These connections are drawn as horizontal lines resembling mountain ladders.

% This design aggregates inheritance trees into coherent mountain groups and reduces link complexity through prefix aggregation. 
% However, curved concatenated contours blur group boundaries, and the internal tree structure is hard to interpret.

% This design aggregates inheritance trees into coherent ``mountain ranges'' and reduces link complexity through prefix aggregation. However, the curved and concatenated contours make boundaries between groups less distinguishable, and the internal tree structure remains difficult to interpret.

\textbf{Design Iteration 3} (Fig.~\ref{fig:Inheriting Mountain View}c).
In the final design, each inheritance group is represented as an individual mountain peak, with painter nodes arranged vertically along its right edge. This makes each group directly readable as a clear vertical unit. Child groups extend horizontally to the right from their parent nodes, so core parent-child relationships can be traced through the rightward growth of the layout. Compared with the previous iteration, this design makes group boundaries within the same tree clearer and the internal tree structure easier to follow. Non-core inheritance links remain encoded as horizontal ladder lines, consistent with the previous design.

% In the final design, each inheritance group is encoded as an individual mountain peak, with painter nodes arranged along the right-side edge. Each group extends rightward from its core parent group's node, so that the spatial positioning itself conveys the core parent connection. This forms a mountain landscape that grows horizontally to the right, making group boundaries and hierarchical relationships within inheritance trees immediately clear. Non-core inheritance links remain as horizontal ladder lines, consistent with the previous design.

When inheritance groups are numerous, a single-row layout aligns all painters on a unified timeline but creates an excessively wide view that limits coordinated analysis. Moreover, later historical periods tend to have more recorded painters, causing dense clusters in the lower region and long ladder connections. To balance timeline comparison and spatial compactness, we provide both single-row and multi-row layouts that users can switch between (Fig.~\ref{fig:Inheriting Mountain View}d,e). In the multi-row layout, the system places connected groups in the same or higher rows to preserve the general top-down inheritance direction. The horizontal order across rows is determined by a bi-directional barycenter method~\cite{liu2013storyflow} to reduce edge crossings. Each row has a secondary timeline on the left. For cross-row connections, links extend from the source group to the inter-row region, bend horizontally, and connect to the target group.

% When the number of inheritance groups is large, a single-row horizontal layout allows all painters to be compared along a unified timeline, but results in an excessively wide view that limits coordinated analysis with other views. Moreover, later historical periods tend to have more recorded painters, causing dense clusters in the lower region and excessively long ladder connections, which reduce readability. To address this, we adopt a multi-row layout strategy. The system divides the view into multiple rows according to inheritance relationships, placing groups with inheritance connections in the same row or higher rows to preserve the general top-down inheritance direction. The horizontal order of inheritance groups across all rows is determined using a bi-directional barycenter method~\cite{liu2013storyflow} to minimize edge crossings. Each row has a secondary timeline on the left indicating its time span. For cross-row connections, links extend vertically from the top of the source group into the inter-row region, bend horizontally toward the target group, and connect vertically to the destination. Users can freely switch between the two layouts to balance timeline comparison and spatial compactness (Fig.~\ref{fig:Inheriting Mountain View}d,e).

Within each inheritance group (mountain peak), color proportions reflect the distribution of painters from different cohorts. Hovering over a painter node or an inheritance group highlights the associated inheritance chains and predecessor nodes.
Mouse scrolling zooms the view; at lower zoom levels, only painters with more disciples are labeled, while higher zoom levels reveal additional details. Users can left-click an inheritance group or a painter node to inspect details in the Detail View; a floating toolbar also appears, providing the \textit{Switch Branch} button to change the core parent group and restructure branches. Users can right-click groups or nodes as rules for the selected cohort; the floating toolbar also provides quick selection tools, such as selecting all descendants of the current group or freely selecting groups.

% the floating toolbar additionally provides the \textit{Select SubGroup} button to quickly select the current group together with all its descendant groups.

The control panel above the view provides two adjustable parameters. \textit{Min Group Size} controls the minimum number of painters required for an inheritance tree to be displayed, with the default value automatically adjusted based on the total number of painters in the view. \textit{Clustering} controls the coverage threshold $\tau$ defined in Section~\ref{Inheritance Forest}, with a default value of $0.65$. A lower value relaxes the requirement for selecting core parent groups and therefore merges inheritance groups more readily into larger trees, while a higher value applies a stricter requirement and results in more separated trees.

% controls how readily inheritance groups are merged into the same tree, ranging from 0.5 to 1. It corresponds to $1 - \tau$, where $\tau$ is the coverage ratio threshold defined in Section~\ref{Inheritance Forest}. A higher value causes groups to cluster more tightly, forming larger trees, while a lower value results in more separated groups.

\textbf{3) Identity, Map, and Timeline Views}

The \textit{Identity View} (Fig.~\ref{fig:System}B3) uses a Treemap to visualize the distribution of painters' identity features. Rectangle sizes encode the proportion of each identity category, and nested rectangles reflect the identity hierarchy. Right-clicking an identity node adds the rule ``painters with this identity'' to the selected cohort, and the corresponding rectangle border is highlighted in the rule's logical color.

The \textit{Map View} (Fig.~\ref{fig:System}B4) presents the spatial distribution of painters across geographic regions, with pie charts overlaid at the center of each region. Chart sizes are proportional to the number of painters. Right-clicking a region adds the rule ``painters born in this region,'' and the region boundary is highlighted in the rule's color.

The \textit{Timeline View} (Fig.~\ref{fig:System}B5) displays the distribution of painters across historical periods. Right-clicking a time period adds the rule ``painters from this period'' to the selected cohort, and the corresponding time block is highlighted accordingly.

All three views support hierarchical exploration via the middle mouse button, enabling users to navigate between coarse and fine granularities. For example, users can drill down from a broad dynasty (e.g., Song Dynasty) to finer historical periods (e.g., Northern Song and Southern Song) in the Timeline View, reveal finer regional divisions in the Map View, or explore specific sub-categories in the Identity View.

\subsection{Detail View}
The Detail View (Fig.~\ref{fig:System}C) provides detailed information about selected cohorts and features, updated through interactions from both the Cohort Management View and the Painter Feature View. In \textit{Cohort Detail} mode (Fig.~\ref{fig:System Interaction} left), the view displays the core features of the selected cohort, along with a complete member list. Core features are determined by a relevance threshold that can be adjusted via a slider, and users can also manually add or remove core features. The member list shows each painter's name, associated feature labels, and a color-coded relevance bar indicating each painter's relevance to the cohort (Section~\ref{Relevance-based Painter Recommendation}), ranging from green (low) to red (high). A relevance threshold slider allows users to quickly filter out low-relevance members for focused inspection. In \textit{Feature Detail} mode (Fig.~\ref{fig:System Interaction} right), the view displays detailed information for the clicked feature. For example, clicking a painter shows their profile, including basic information, feature labels, inheritance chains, and artistic-style descriptions with extracted labels highlighted; clicking an artistic feature shows its description and the associated painters.
\section{{Evaluation}}
\label{sec:evaluate}

We evaluated HPC-Vis through case studies with domain experts, a user study comparing with the traditional workflow, and technical evaluations of the inheritance reconstruction algorithm and visualization. A dynamic demonstration of these case studies is provided in the supplementary video.

% 案例 1：构建花鸟画家群体
% 专家A旨在探索中国花鸟画家的群体。在艺术标签视图中，他使用OR规则选择了“花鸟”主题标签，将所有具有此功能的画家添加到队列中。通过观察技术分布，他发现Xieyi和Gongbi技术都占主导地位，并点击每种技术以查看详细视图中的描述。由于这两种技巧代表了截然不同的风格方法，他假设花鸟画家可能形成两个不同的风格阵营，值得进一步比较
\subsection{Case Studies}
We present two progressive case studies by domain expert E1, showing how HPC-Vis supports cohort exploration from macro-level comparison to fine-grained internal analysis.
% We present two progressive case studies conducted by domain expert E1, illustrating how HPC-Vis supports cohort exploration from macro-level comparison to fine-grained internal analysis.
% \subsection{Case Study: Exploring Flower-Bird Painter Cohorts}
\subsubsection{Case 1: Exploring Flower-Bird Painter Cohorts}
Expert E1 aimed to explore the cohort of Chinese flower-bird painters. In the \textit{Artistic Label View}, he selected the ``Flower-Bird'' subject label using an OR rule, forming the \textcolor{cohortgreen}{$\blacksquare$} Flower-Bird Cohort. Observing the technique distribution, he found that both ``Xieyi'' and ``Gongbi'' techniques were dominant and clicked on each to inspect their descriptions in the \textit{Detail View}. As these two techniques represent contrasting stylistic approaches, he hypothesized that flower-bird painters may form two distinct stylistic camps worth further comparison.

\begin{figure}[!t]
    \setlength{\belowcaptionskip}{-0.3cm}
    \centering 
    \includegraphics[width=0.95\columnwidth]{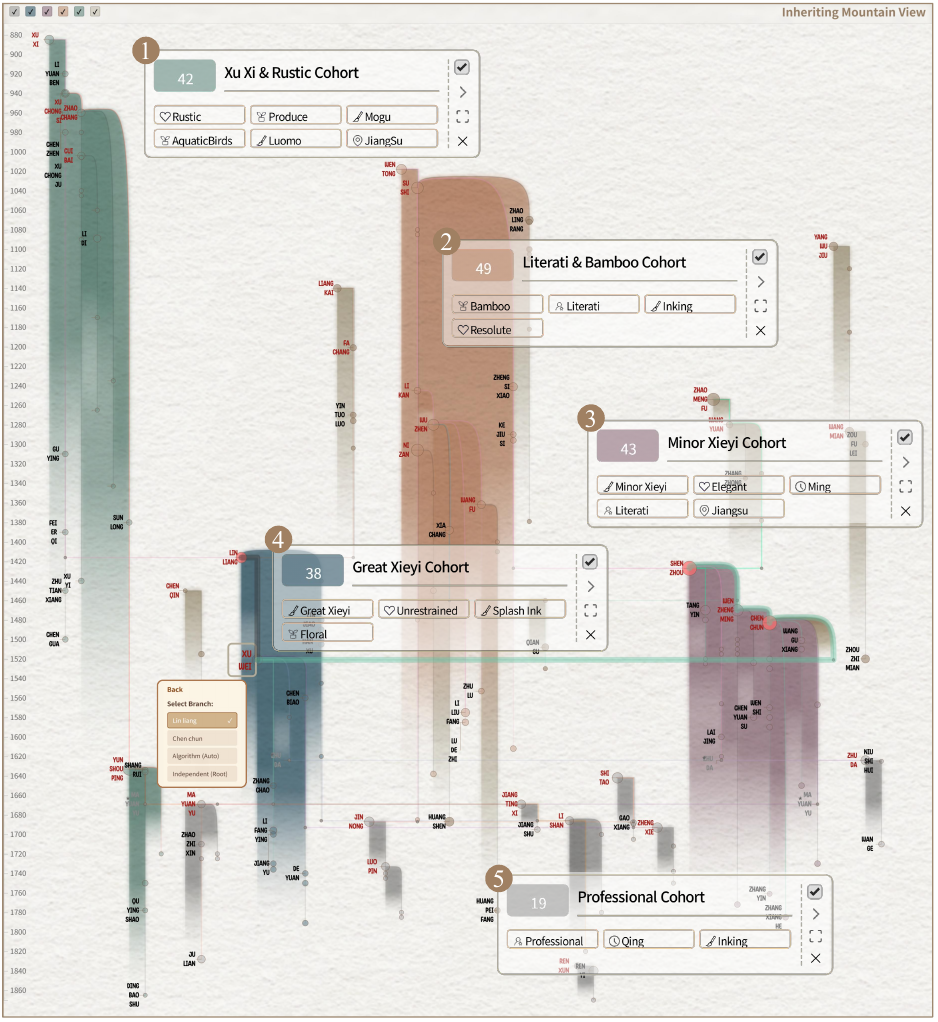}
    \caption{The \textit{Inheriting Mountain View} after entering the ``Xieyi Flower-Bird Cohort'' shows the inheritance structures within this cohort. Five sub-cohorts identified by E1 are annotated with their corresponding cohort cards near the respective mountains.}
    \label{fig:Case1}
\end{figure}

% 构建两个子群体：写意花鸟画家群体和工笔花鸟画家群体
% E1 then applied an AND rule to select ``Xieyi'' painters within the cohort, creating the \textcolor{cohortgreen}{$\blacksquare$} Xieyi Flower-Bird Cohort. He constructed another cohort by combining ``Flower-Bird'' and ``Gongbi'' with AND rules, forming the \textcolor{cohortred}{$\blacksquare$} Gongbi Flower-Bird Cohort, yielding two candidate cohorts for contrastive analysis (Fig.~\ref{fig:System}).

E1 then applied an AND rule to select ``Xieyi'' painters within the cohort, creating the \textcolor{cohortgreen}{$\blacksquare$} Xieyi Flower-Bird Cohort. He also combined ``Flower-Bird'' and ``Gongbi'' with AND rules to form the \textcolor{cohortred}{$\blacksquare$} Gongbi Flower-Bird Cohort, yielding two candidate cohorts for contrastive analysis (Fig.~\ref{fig:System}).

% 多视角观察群体特征
To validate this hypothesis, E1 examined both cohorts from multiple perspectives. In the \textit{Artistic Label View} (Fig.~\ref{fig:System}B1), he used focus expansion on ``Flower-Bird'' to reveal second-level subject labels, and context expansion on ``Floral'' and ``Birds'' to observe finer distributions. The ``Gongbi'' cohort frequently depicted exotic birds and rare flowers (e.g., peacocks, cranes, peonies), whereas the ``Xieyi'' cohort focused on natural subjects (e.g., plum, orchid, bamboo, chrysanthemum, and fruits). The \textit{Identity View} (Fig.~\ref{fig:System}B3) further reinforced this divergence: ``Gongbi'' painters were largely Art Academy painters, while ``Xieyi'' painters were predominantly literati. The \textit{Map View} (Fig.~\ref{fig:System}B4) showed that ``Gongbi'' painters were concentrated in northern regions around historical capitals, while ``Xieyi'' painters were based in southern provinces. The \textit{Inheriting Mountain View} (Fig.~\ref{fig:System}B2) revealed that the two cohorts were largely distributed across different mountain clusters. These multi-dimensional observations confirmed the overall divergence between the two cohorts.

\subsubsection{Case 2: Exploring Xieyi Flower-Bird Sub-cohorts}
To conduct a more detailed analysis of the ``Xieyi'' cohort, E1 entered the ``Xieyi Flower-Bird Cohort'' through the \textit{Cohort Management View}. The \textit{Painter Feature View} now displayed the painter feature distributions exclusively within this cohort. In the \textit{Inheriting Mountain View}, the minimum mountain scale automatically reduced, revealing smaller mountain clusters and isolated painters. By exploring the inheritance trees in the \textit{Inheriting Mountain View} and coordinating with the \textit{Artistic Label View} and \textit{Detail View}, E1 progressively identified and defined five sub-cohorts (Fig.~\ref{fig:Case1}):

% 六个子群体详细介绍
\textbf{\textcolor{cohortgreen}{$\blacksquare$} Xu Xi's Rustic Cohort.} 
E1 right-clicked the Xu Xi mountain (Fig.~\ref{fig:Case1}\ding{192}) in the \textit{Inheriting Mountain View} to add it to the initial cohort. In the \textit{Artistic Label View}, the cohort showed a predominantly ``Rustic'' emotion style, and expanding the ``Flower-Bird'' subject label revealed a focus on natural objects (e.g., wildflowers, aquatic birds, garden vegetables). The technique distribution was diverse, with ``Luomo,'' ``Mogu,'' and ``Gongbi'' all showing relatively high proportions. Looking further into the circle pack, a combination of ``Mogu'' and ``Gongbi'' was present, while ``Luomo'' appeared independently without forming combinations with other techniques. 

Given the mountain's long time span, E1 hypothesized that these techniques might represent different developmental stages of the lineage.
To verify this hypothesis, E1 added ``Luomo'' as an AND rule, narrowing the cohort to painters proficient in this technique. In the \textit{Inheriting Mountain View}, these painters were concentrated in the early branches directly inheriting from Xu Xi. He then removed this rule and added ``Mogu'' as an AND rule instead. The resulting painters first appeared in the Cui Bai sub-branch, decreased during the middle period, and re-emerged in the Qing Dynasty with a notable sub-branch.

By inspecting key painter nodes at branch points in the \textit{Detail View}, E1 confirmed this hypothesis: the lineage originated with Xu Xi's ``Luomo'' technique, which gradually gave way to ``Mogu'' combined with ``Gongbi'', was ultimately revitalized by Yun Shouping and his disciples in the Qing Dynasty.

% 文人墨竹群体
\textbf{\textcolor{cohortred}{$\blacksquare$} Literati Bamboo Cohort.} E1 created a new cohort and added the mountain led by Wen Tong and Su Shi (Fig.~\ref{fig:Case1}\ding{193}) to the cohort rules. The system automatically recommended core features for this cohort in the Cohort List, including ``Bamboo'' subject, ``Resolute'' emotion, ``Ink Wash'' technique, and ``Literati'' identity. These recommendations allowed E1 to quickly locate and verify the relevant features across views. He further inspected the cohort member list in the \textit{Detail View} and filtered out less relevant painters to finalize the cohort.

%  小写意和大写意群体：子群体区分与传承链调整
\textbf{\textcolor{cohortpurple}{$\blacksquare$} Minor Xieyi and \textcolor{cohortblue}{$\blacksquare$} Great Xieyi Cohorts.} 
While exploring the inheritance structure of the Shen Zhou mountain, E1 noticed that the Xu Wei group had a non-core inheritance link connecting to the Lin Liang mountain. By inspecting their feature information in the \textit{Detail View} (Fig.~\ref{fig:System Interaction}C), E1 found that Xu Wei and his predecessor Lin Liang both exhibited ``Great Xieyi'' technique with ``Unrestrained'' emotion, while his other predecessors Shen Zhou and Chen Chun were predominantly characterized by ``Minor Xieyi'' technique with ``Elegant'' emotion. This suggested that Xu Wei's artistic features may be more closely aligned with Lin Liang's tradition.

To verify this hypothesis, E1 created two cohorts for comparison (Fig.~\ref{fig:System Interaction}B): one including the branches from Shen Zhou to Chen Chun, and another including the Xu Wei branch. By highlighting their core features, the \textit{Artistic Label View} automatically expanded the ``Xieyi'' technique label, clearly showing the ``Minor Xieyi'' versus ``Great Xieyi'' divergence. Based on this observation, E1 used the \textit{Switch Branch} function to reassign Xu Wei's core inheritance link to the Lin Liang mountain, and added the Lin Liang branch to the Xu Wei cohort. The two cohorts were then named as the Minor Xieyi Cohort (Fig.~\ref{fig:Case1}\ding{194}) and the Great Xieyi Cohort (Fig.~\ref{fig:Case1}\ding{195}) respectively.

% 职业画家群体：小规模零散群体，推荐扩展
\textbf{\textcolor{cohortgray}{$\blacksquare$} Professional Painter Cohort.} E1 noticed several small and scattered mountains in the late Ming to early Qing period. By clicking on painter nodes such as Jin Nong and Li Shan and inspecting their features in the \textit{Detail View}, he found that they exhibited diverse artistic features but shared a common professional painter identity. He created a new cohort and added these two painters to the cohort rules. He then increased the weights of ``Identity'' and ``Time'' dimensions in the Recommendation panel and triggered the recommendation. The system generated a recommended cohort, whose feature distributions could be examined across the feature views for verification. After inspecting the results, E1 merged the recommended painters into the cohort to form the Professional Painter Cohort (Fig.~\ref{fig:Case1}\ding{196}).

% 总结 HPC-Vis
After completing the exploration, E1 commented that HPC-Vis accelerated the iterative cycle of ``formulate hypotheses---define cohorts---validate through analysis.'' The \textit{Inheriting Mountain View} provided an intuitive lineage overview, multi-view coordination enabled rapid cross-dimensional validation, and the recommendation function offered effective computational guidance for cohort analysis. Overall, the system enabled progressive exploration from macro- to micro-level distributions, improving efficiency.

% After completing the exploration, E1 commented that HPC-Vis significantly accelerated the iterative cycle of ``formulate hypotheses---define cohorts---validate through analysis.'' The \textit{Inheriting Mountain View} provided an intuitive overview of inheritance lineages, while the multi-view coordination allowed rapid cross-validation of cohort features across dimensions. The recommendation function offered effective computational guidance for cohort analysis, including core feature identification, cohort refinement, and expansion. Overall, the system enabled a progressive exploration from macro-level to micro-level distributions, significantly improving the efficiency.

\subsection{User Study}
We evaluated HPC-Vis through a user study and interviews with collaborating domain experts (E1--E4).

% To evaluate the effectiveness and practicality of HPC-Vis, we conducted a user study and performed interviews with our collaborating domain experts (E1--E4) to gather their feedback.

\textbf{Participants.} We recruited 24 participants, including the 4 collaborating experts, art history graduate students, and amateur enthusiasts of Chinese painting history.
% We recruited 24 volunteers to participate in our study. Among them, 4 are our collaborating domain experts, while the others include graduate students in art history and amateur enthusiasts of Chinese painting history.

\begin{figure}[!t]
    \setlength{\belowcaptionskip}{-0.3cm}
    \centering 
    \includegraphics[width=0.95\columnwidth]{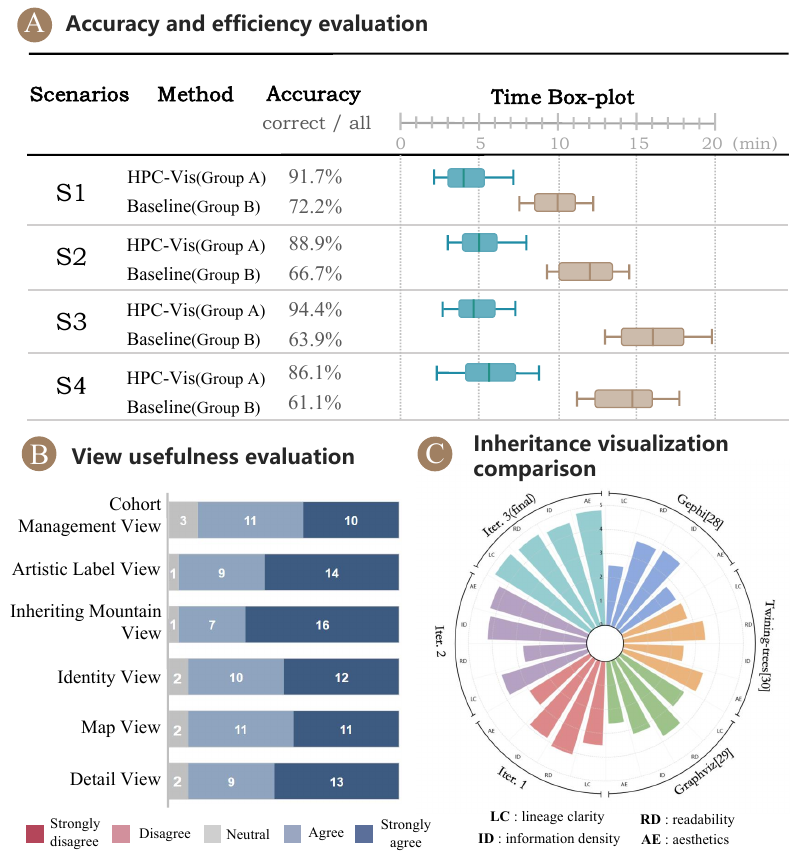}
\caption{Evaluation results. (A) Accuracy and efficiency evaluation. (B) View usefulness evaluation. (C) Inheritance visualization comparison.}

    \label{fig:study}
\end{figure}

\textbf{Study Procedure.} 
Participants first received a system demonstration and then freely explored HPC-Vis for 10 minutes.
% Participants first received a demonstration of the system interface and workflow, followed by a 30-minute session to familiarize themselves with the system. 
We designed tasks covering four types of analytical scenarios commonly encountered in painter cohort research:

\textbf{S1.} Given a set of painters, identify their major shared features across multiple dimensions.

\textbf{S2.} Given two painter cohorts, compare their differences across multiple dimensions.

\textbf{S3.} Given a set of painters with complex inheritance relationships, identify the main lineage structure.

\textbf{S4.} Given a cohort and a list of candidate painters, determine which candidates should be included in the cohort.

Each scenario contains 4 tasks with definitive answers. The task list is provided in the supplementary material. We recorded the accuracy and completion time for each task throughout the study. The study consisted of two phases. In Phase 1, all participants completed 4 tasks (one per scenario) using HPC-Vis. Based on Phase 1 performance, participants were divided into two groups of balanced system proficiency: Group A and Group B, each consisting of 12 members with domain experts evenly distributed. In Phase 2, both groups completed the remaining 12 tasks (three per scenario) on the same set of questions. Group A used HPC-Vis, while Group B followed the traditional workflow, with painter data provided as spreadsheets and the inheritance network rendered in Gephi~\cite{Gephi}. Upon completion, participants rated each view's usefulness on a 5-point Likert scale and provided open-ended comments.

% Upon completion, all participants filled out a questionnaire rating each view on a 5-point Likert scale based on overall usefulness for cohort analysis. Participants were also invited to provide open-ended comments.

\textbf{Results.} As shown in Fig.~\ref{fig:study}(A), Group A (HPC-Vis) consistently outperformed Group B (traditional workflow) across all four task scenarios in both accuracy and completion time. Participants in Group B remarked that the traditional workflow felt considerably slower and more error-prone compared to their Phase 1 experience with the system, confirming the practical value of HPC-Vis. As shown in Fig.~\ref{fig:study}(B), all views received positive ratings. Among them, the \textit{Inheriting Mountain View} received the highest rating, with participants noting that it provided an intuitive overview of inheritance structures, making complex lineage relationships much easier to understand.

\textbf{Expert feedback.} We conducted in-depth interviews with E1--E4 after the study, and summarize their observations:

1) Overall Workflow. All experts agreed that HPC-Vis substantially improved their research efficiency. They appreciated that the system preserved their familiar research approach while providing computational and visual support at each step. E1 commented that the multi-view coordination allowed him to ``observe cohort features from different perspectives simultaneously,'' enabling rapid analysis of cohort features that would otherwise require extensive manual cross-referencing. E3 highlighted the recommendation function for its ability to quickly identify core features and discover overlooked painters. E4 noted that the cohort management mechanism, particularly the ability to nest and compare multiple cohorts, greatly facilitated sustained and systematic exploration. 

2) Visualization and Interaction. Experts recognized the effectiveness of the visual designs. The \textit{Inheriting Mountain View} received the strongest praise, with E1 noting that it transformed lineage tracing from a laborious manual task into an intuitive visual exploration. E2 appreciated the \textit{Artistic Label View} for its multi-scale exploration capability, which enabled flexible switching between coarse and fine-grained feature distributions.
Experts also praised the interaction design for being consistent and easy to learn across views.

3) Suggestions. E1 suggested integrating painting images into the system, which would allow users to directly compare visual styles alongside the current text-based features. E3 noted that the analysis quality depends on the underlying data, and incorporating additional data sources could broaden the scope.

\subsection{Technical Evaluation}

We evaluated our core technical contributions, the inheritance reconstruction algorithm and the Inheriting Mountain View visualization, from three perspectives: structural simplification, grouping quality, and visual effectiveness.

\textbf{Structural Simplification.} The original inheritance network contains 2,212 painter nodes and approximately 3,400 edges. After reconstruction, these are reduced to approximately 1,060 inheritance groups and 1,180 inter-group connections, achieving roughly 52\% reduction in nodes and 65\% reduction in edges while preserving essential inheritance information.

% \textbf{Grouping Quality.} To assess whether the reconstructed inheritance trees align with art-historical knowledge, we compare our method against several widely used graph clustering algorithms. Using well-documented painting schools recognized by our collaborating historians as ground truth (e.g., the Wu School, the Zhe School), we measure \textit{coverage} (the proportion of known school members contained within a single group) and \textit{purity} (the proportion of members within a group that belong to the same school). As shown in Table~\ref{tab:grouping}, our method achieves higher scores on both metrics. Our method is specifically designed for inheritance networks, leveraging both the directionality of artistic inheritance and the continuity of multi-generational lineages, while general clustering algorithms are designed for general graph structures. \

\textbf{Grouping Quality.} To assess whether the reconstructed inheritance trees align with art-historical knowledge, we compare our method with representative baselines including a strong general-purpose community detection method (Leiden), a recent directed graph clustering method (Directed HOSC), and a DAG-specific community detection method (DAG-CD). Using well-documented painting schools recognized by our collaborating historians as ground truth (e.g., the Wu School and the Zhe School), we measure \textit{coverage} (the proportion of known school members contained within a single group) and \textit{purity} (the proportion of members within a group that belong to the same school). As shown in Table~\ref{tab:grouping}, our method achieves the best results on both metrics.
% Our method is specifically designed for inheritance networks, leveraging both the directionality of artistic inheritance and the continuity of multi-generational lineages, whereas the baselines are general graph clustering methods or only partially exploit the DAG structure.

\begin{table}[!t]
    \renewcommand{\arraystretch}{0.95}
    \centering
    \small
    \refstepcounter{table}
    \textsc{TABLE \thetable}\quad Grouping Quality Comparison
    \label{tab:grouping}
    \vspace{0.2em}

    \resizebox{\columnwidth}{!}{
    \begin{tabular}{ccccc}
        \toprule
        Methods
        & \makecell{Leiden\\\cite{traag2019louvain}}
        & \makecell{Directed HOSC\\\cite{laenen2020higher}}
        & \makecell{DAG-CD\\\cite{speidel2015community}}
        & Ours \\
        \midrule
        Coverage & 0.65 & 0.62 & 0.78 & \textbf{0.89} \\
        Purity   & 0.53 & 0.71 & 0.66 & \textbf{0.85} \\
        \bottomrule
    \end{tabular}
    }
\end{table}

% \begin{table}[!t]
%     \renewcommand{\arraystretch}{0.95}
%     \centering
%     \small
%     \refstepcounter{table}
%     \textsc{TABLE \thetable}\quad Grouping Quality Comparison
%     \label{tab:grouping}
%     \vspace{0.2em}

%     \resizebox{\columnwidth}{!}{
%     \begin{tabular}{cccccc}
%         \toprule
%         Methods
%         & \makecell{Louvain\\\cite{blondel2008fast}}
%         & \makecell{Label Propagation\\\cite{raghavan2007near}}
%         & \makecell{Girvan Newman\\\cite{girvan2002community}}
%         & \makecell{Spectral Clustering\\\cite{ng2001spectral}}
%         & Ours \\
%         \midrule
%         Coverage & 0.72 & 0.65 & 0.70 & 0.67 & \textbf{0.89} \\
%         Purity   & 0.68 & 0.63 & 0.71 & 0.66 & \textbf{0.85} \\
%         \bottomrule
%     \end{tabular}
%     }
% \end{table}

% \begin{table}[!t]
%     \renewcommand{\arraystretch}{0.95}
%     \centering
%     \small
%     \caption{Grouping Quality Comparison}
%     \label{tab:grouping}
%     \resizebox{\columnwidth}{!}{
%     \begin{tabular}{cccccc}
%         \toprule
%         Methods
%         & \makecell{Louvain\\\cite{blondel2008fast}}
%         & \makecell{Label Propagation\\\cite{raghavan2007near}}
%         & \makecell{Girvan Newman\\\cite{girvan2002community}}
%         & \makecell{Spectral Clustering\\\cite{ng2001spectral}}
%         & Ours \\
%         \midrule
%         Coverage & 0.72 & 0.65 & 0.70 & 0.67 & \textbf{0.89} \\
%         Purity   & 0.68 & 0.63 & 0.71 & 0.66 & \textbf{0.85} \\
%         \bottomrule
%     \end{tabular}
%     }
% \end{table}

\textbf{Visual Effectiveness.} We invited domain experts and visualization researchers to compare the Inheriting Mountain View against several alternative visualization methods (shown in the supplementary material), including existing tools and our earlier design iterations, applied to the same 60-painter subset. Participants rated each method on four criteria using a 5-point Likert scale: \textit{lineage clarity} (how clearly inheritance lineages can be identified), \textit{readability} (how easily the overall structure can be read and understood), \textit{information density} (how much effective information is conveyed without causing visual clutter), and \textit{aesthetics} (how visually appealing the representation is). As shown in Fig.~\ref{fig:study}(C), our method received the highest ratings across all criteria.
\section{{Conclusion}}

In this paper, we presented HPC-Vis, a visual analytics system for interactive exploration of historical painter cohorts. Through close collaboration with domain experts, we identified key challenges in traditional painter cohort research and proposed an improved workflow integrating structured feature construction, visualization-assisted exploration, algorithm-based recommendation, and unified cohort management. We developed an inheritance reconstruction algorithm that transforms complex multi-parent networks into clear forest structures, and designed the Inheriting Mountain View to intuitively reveal lineage relationships. The case study, user study, and technical evaluations demonstrate the effectiveness of the proposed approach. While HPC-Vis currently focuses on Chinese historical painters, the analytical workflow and core techniques are potentially generalizable to other domains where cohort analysis involves multi-dimensional features and inheritance-like relationships, such as calligraphers, academic lineages, and traditional Chinese medicine cohorts.

\textbf{Limitations and Future Work.} The quality of the analysis is inherently dependent on the accuracy and completeness of the underlying data. The current dataset covers relatively well-known painters, and historical records inevitably contain gaps and biases. We plan to incorporate multiple data sources to broaden the coverage, though this will also require addressing challenges of cross-database integration and consistency. %
The artistic feature extraction currently relies on LLM with limited human verification. Developing a human-in-the-loop approach for collaborative multi-level concept extraction from large-scale textual data will allow domain experts to more efficiently review and refine the extracted features. %
%Beyond text-based features, we also plan to integrate painting images to enable direct visual style comparison. 
% Furthermore, the current system separates different feature dimensions into independent views to avoid visual interference. However, we observed that experts particularly enjoyed exploring cohorts along inheritance relationships, tracing how features in other dimensions evolved through these lineages. We therefore plan to further refine the visual design of the \textit{Inheriting Mountain View} to better support this lineage-driven exploration pattern.

\bibliographystyle{IEEEtran}

\bibliography{references}

\AuthorBio[5]{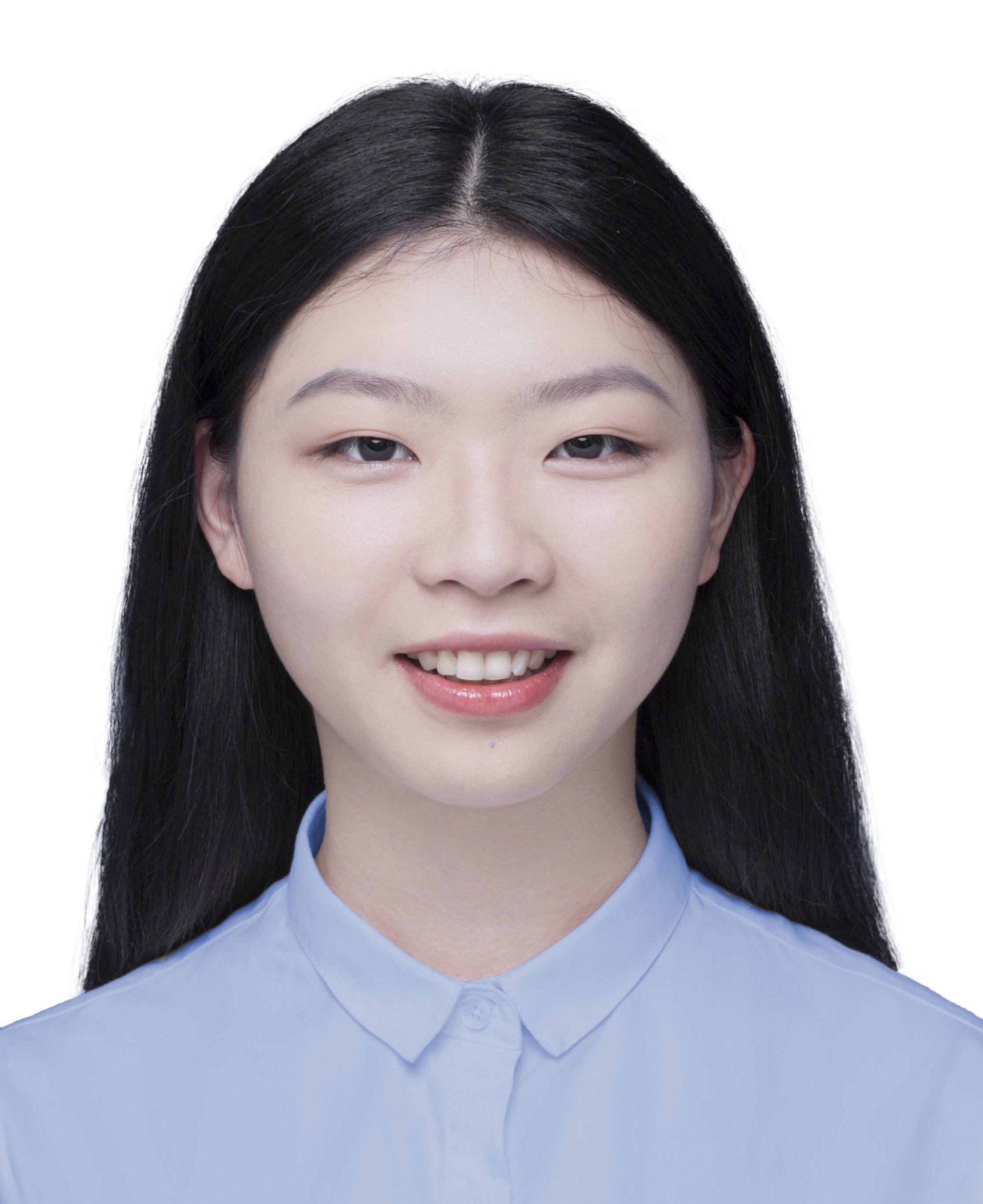}{Yingping Yang}{ received the B.S. degree from Zhejiang University of Technology, China, in 2023. She is pursuing the Ph.D. degree in Software Engineering at Zhejiang University of Technology, China. Her research interests include visual analytics, cultural computing and human-computer interaction.}

\AuthorBio{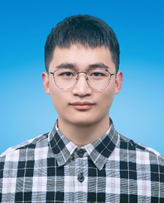}{Guangtao You}{ received the B.S. degree in Software Engineering from Zhejiang University of Water Resources and Electric Power, China, is currently pursuing the M.S. degree in Software Engineering at Zhejiang University of Technology. His research interests include human-computer interaction.
}

\AuthorBio{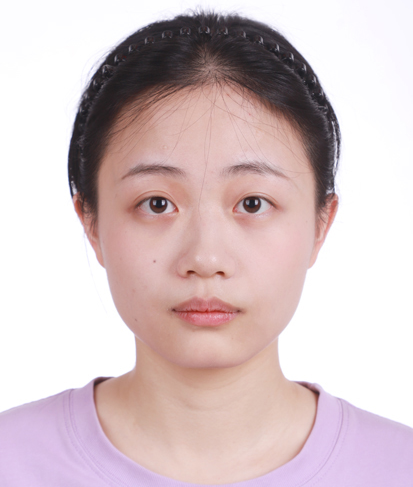}{Wenwen Li}{ is currently pursuing the B.S. degree in Computer Science and Technology at Zhejiang University of Technology, China. Her research interests include visualization and digital humanities.
}

\AuthorBio{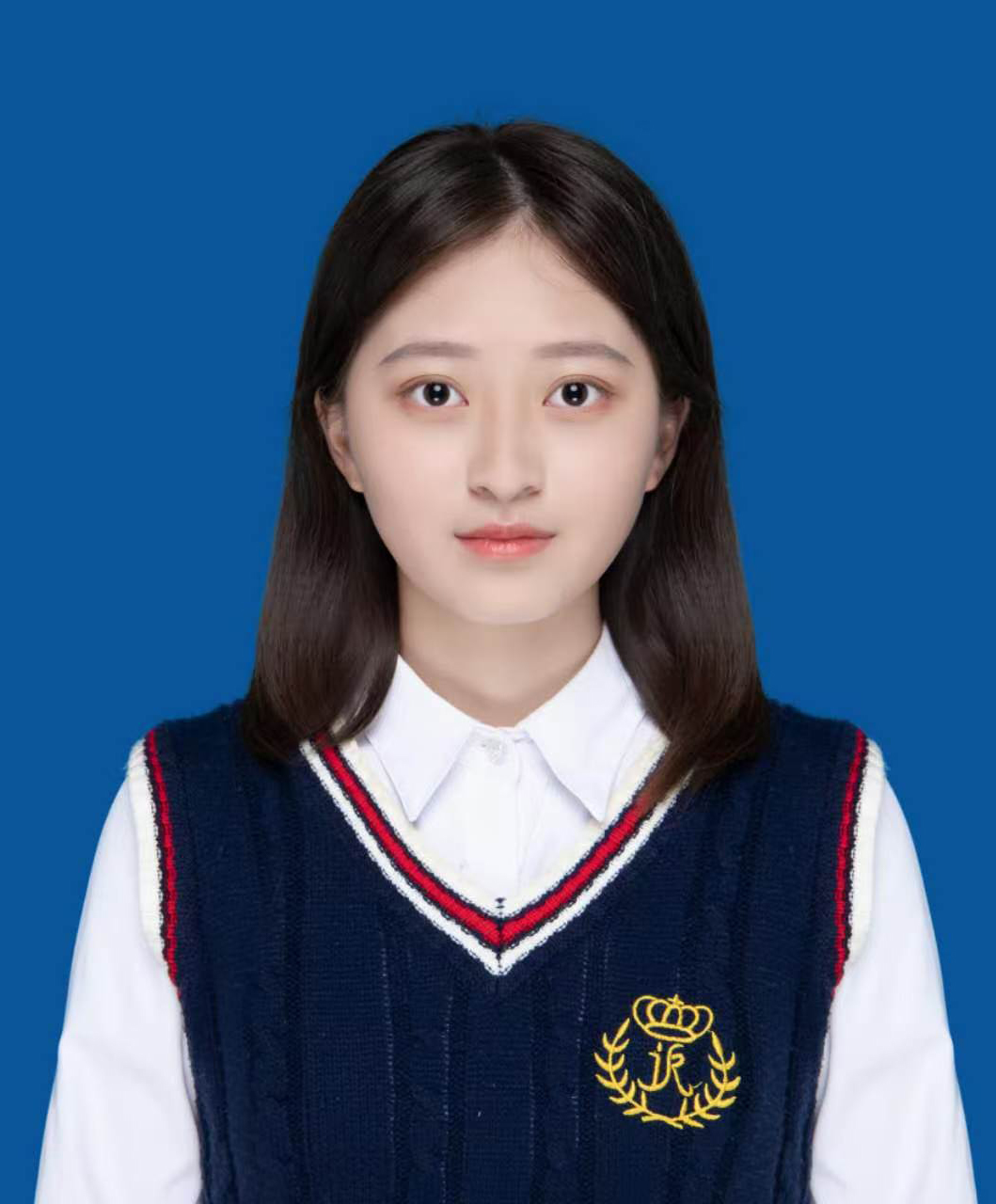}{Jiayi Chen}{ is currently pursuing the B.S. degree in Digital Media Technology at Zhejiang University of Technology, China. Her research interests include visual analytics and geospatial visualization.
}

\AuthorBio{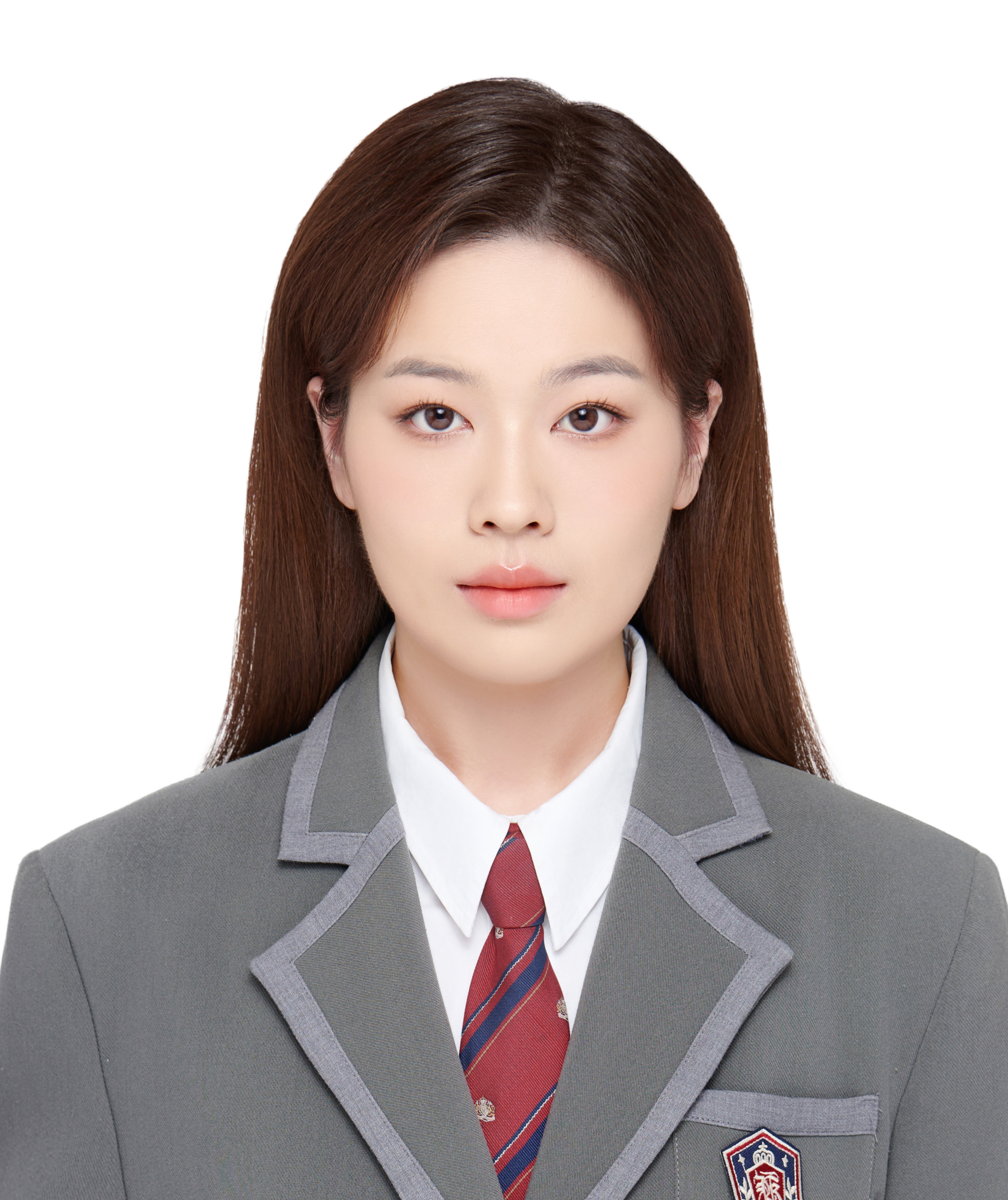}{Yumeng Zhang}{ received the B.S. degree from Zhengzhou University of Light Industry, China. She is pursuing the M.S. degree in Software Engineering at Zhejiang University of Technology. Her research interests include digital humanities and visual analytics.
}

\AuthorBio{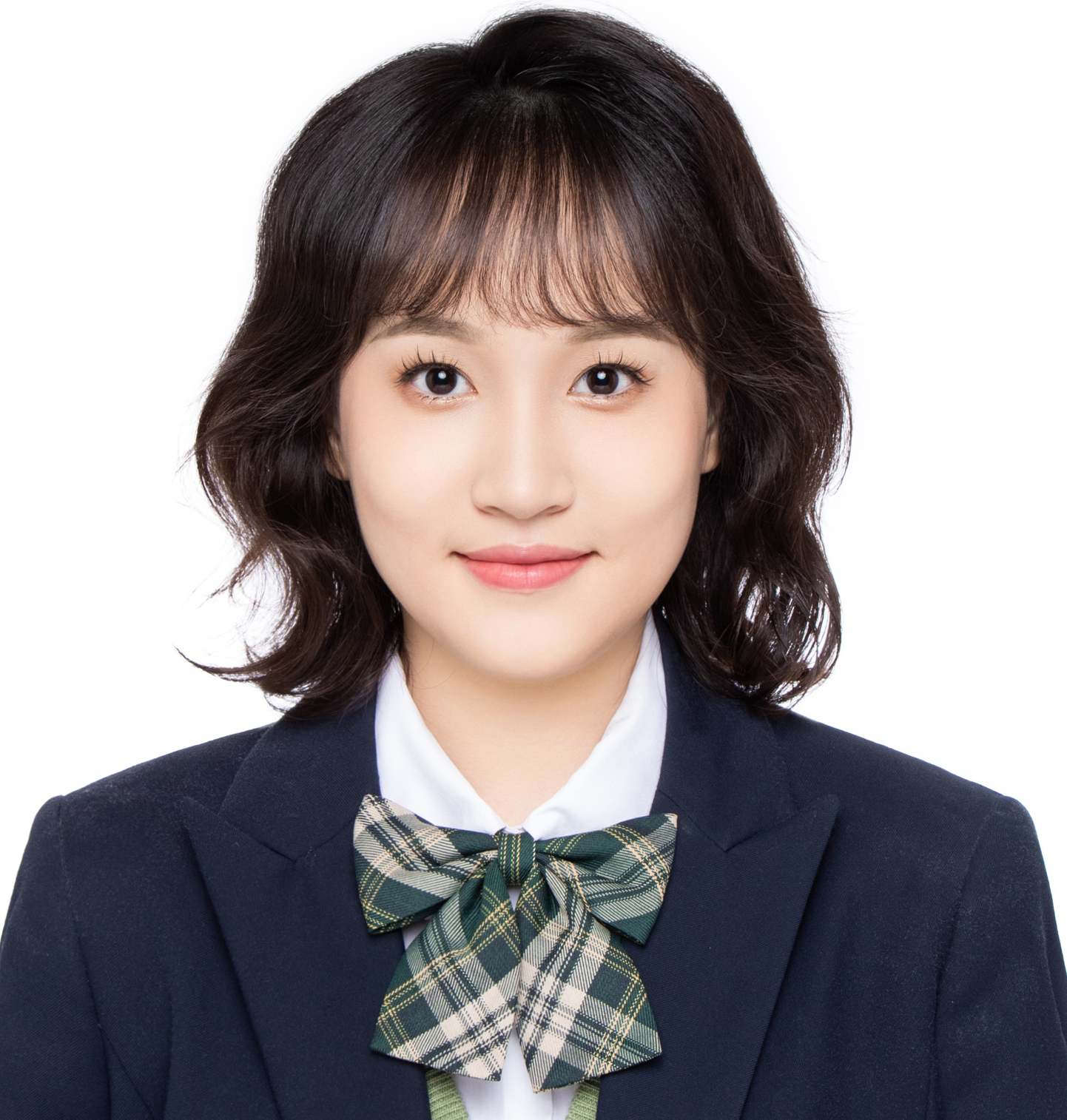}{Yuxin Lei}{ is currently pursuing the B.S. degree in Digital Media Technology at Zhejiang University of Technology, China. Her research interests include visual analytics and digital humanities.
}

\AuthorBio{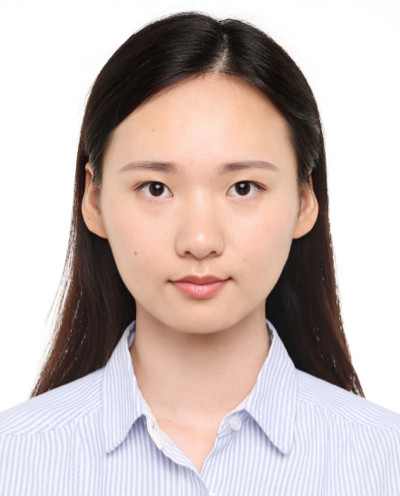}{Wei Zhang}{ received her Ph.D. degree in Design Science from Zhejiang University in 2025. She is currently a lecturer with the School of Computer and Computing Science, Hangzhou City University. Her research interests include visual analytics, cultural computing and the application of AI techniques for the interpretation and understanding of art and cultural heritage data.
}

\AuthorBio{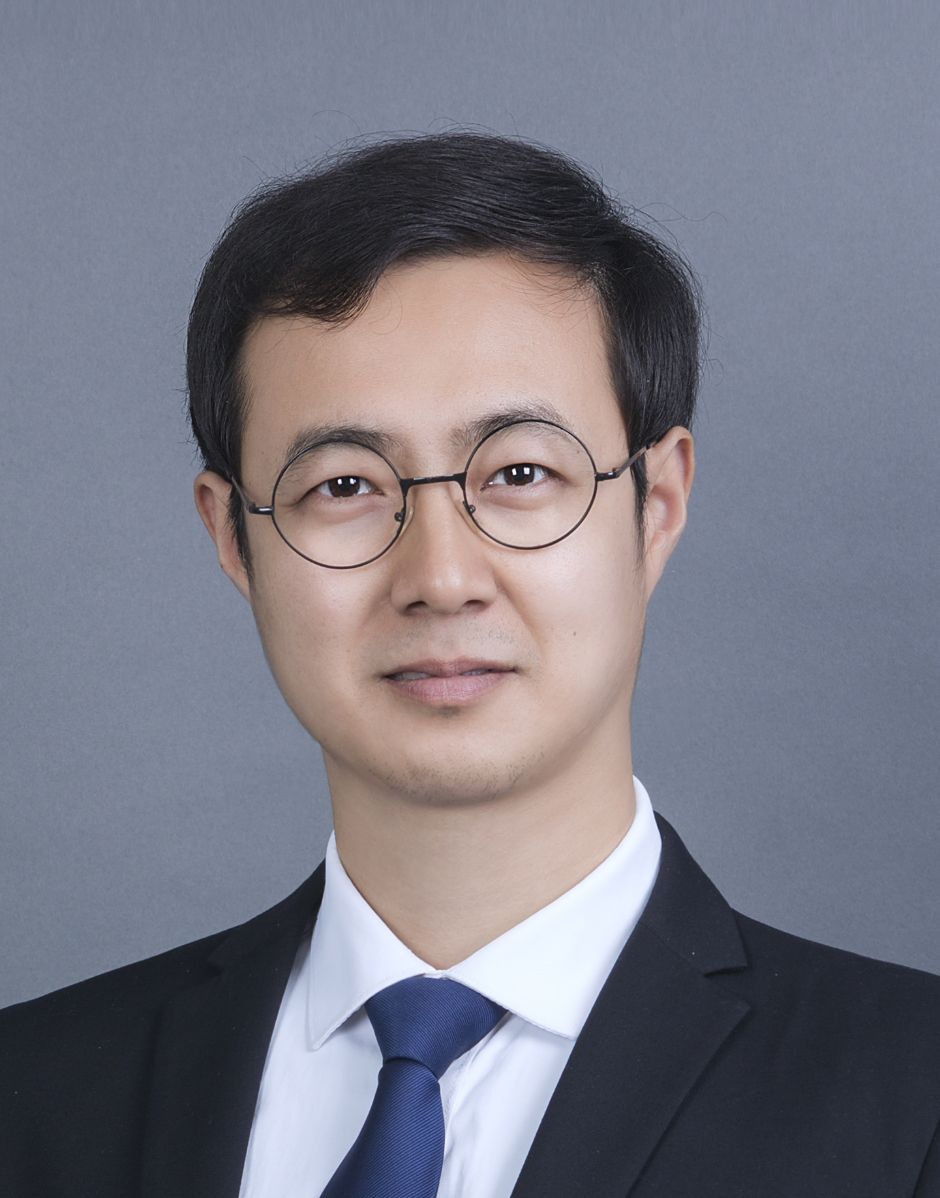}{Jiazhou Chen}{ is a professor with the School of Computer Science and Technology in Zhejiang University of Technology, the deputy director of the Computer Software Institute. He received his double Ph.D. degrees in Bordeaux University, France, and the State Key Laboratory of CAD\&CG in Zhejiang University, China. His research interests include computer graphics, visual analytics, and digital humanities.
}

\AuthorBio{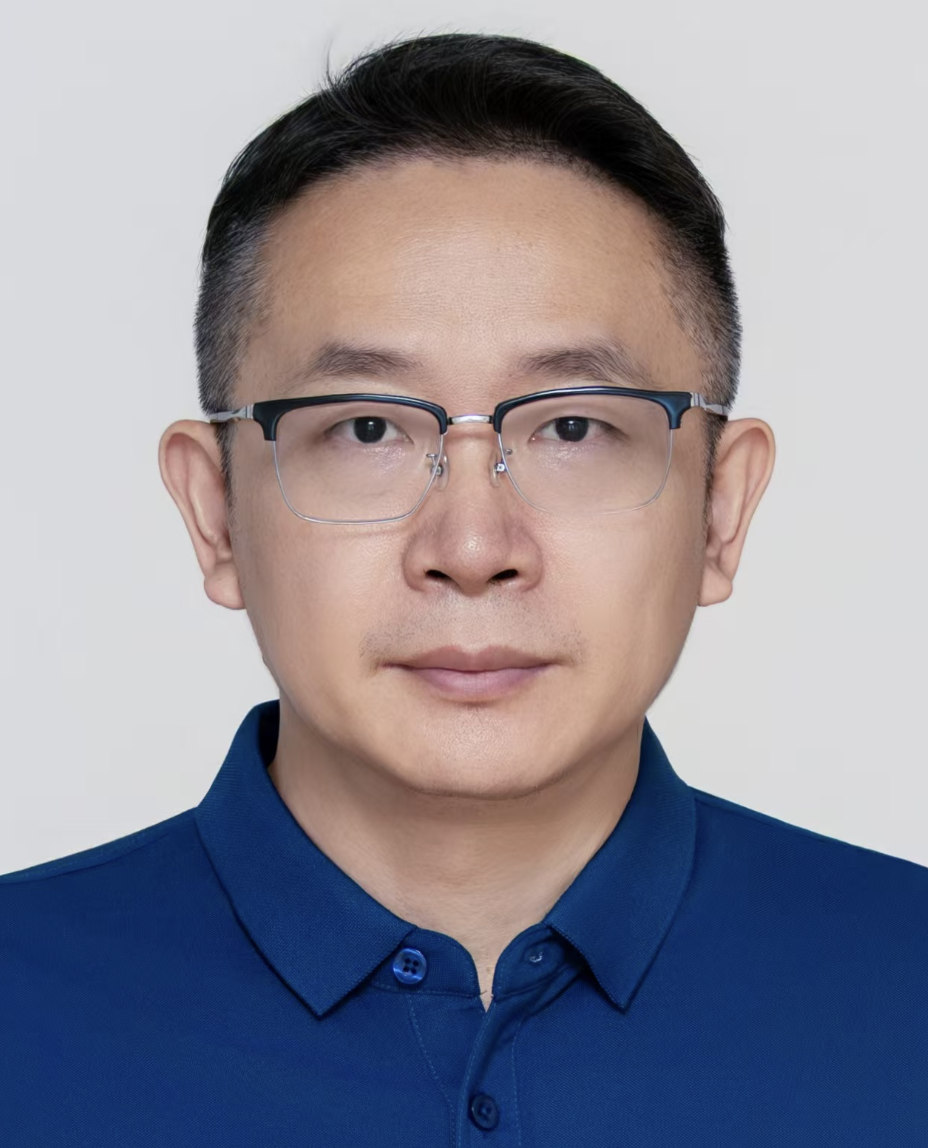}{Wei Chen}{ is a professor with the State Key Lab of CAD\&CG, Zhejiang University. His research interests include visualization and visual analysis, and has published more than 70 IEEE/ACM Transactions and IEEE VIS papers. He actively served as guest or associate editors of the ACM Transactions on Intelligent System and Technology, the IEEE Computer Graphics and Applications, and Journal of Visualization.
}

\end{document}